\documentclass[final]{svjour2}
\usepackage{graphicx}
\usepackage{rotating}
\usepackage{amssymb}
\usepackage{mathptmx}
\usepackage[numbers]{natbib}
\makeatletter \journalname{Journal of Low Temperature Physics}
\bibpunct{}{}{,}{s}{}{,}

\begin{document}

\newcommand{\hdblarrow}{H\makebox[0.9ex][l]{$\downdownarrows$}-}
\title{Thermodynamic Measurements in a\\ Strongly Interacting Fermi Gas}

\author{Le Luo$^{1,2}$ \and J. E. Thomas$^{1,3}$}

\institute{1:Department of Physics, Duke University, Durham, NC
27708
\\2: Current address: Joint Quantum Institute, University of Maryland and National Institute of Standards and Technology, College Park, MD 20742
\\3: \email{jet@phy.duke.edu}}

\date{11.4.2008}

\maketitle


\begin{abstract}
Strongly interacting Fermi gases provide a clean and controllable
laboratory system for modeling strong interparticle interactions
between fermions in nature, from high temperature superconductors
to neutron matter and quark-gluon plasmas. Model-independent
thermodynamic measurements, which do not require theoretical
models for calibrations,  are very important for exploring this
important system experimentally, as they enable direct tests of
predictions based on the best current non-perturbative many-body
theories. At Duke University, we use all-optical methods to
produce a strongly interacting Fermi gas of spin-1/2-up and
spin-1/2-down $^6$Li atoms that is magnetically tuned near a
collisional (Feshbach) resonance. We conduct a series of
measurements on the thermodynamic properties of this unique
quantum gas, including the energy $E$, entropy $S$, and sound
velocity $c$. Our model-independent measurements of $E$ and $S$
enable a precision study of the finite temperature thermodynamics.
The $E(S)$ data are directly compared to several recent
predictions. The temperature in both the superfluid and normal
fluid regime is obtained from the fundamental thermodynamic
relation $T=\partial E/\partial S$ by parameterizing the $E(S)$
data using two different power laws that are joined with
continuous $E$ and $T$ at a certain entropy $S_c$, where the fit
is optimized. We observe a significant change in the scaling of
$E$ with $S$ above and below $S_c$. Taking the fitted value of
$S_c$ as an estimate of the critical entropy for a
superfluid-normal fluid phase transition in the strongly
interacting Fermi gas, we estimate the critical parameters.  Our
$E(S)$ data are also used to experimentally calibrate the endpoint
temperatures obtained for adiabatic sweeps of the magnetic field
between the ideal and strongly interacting regimes. This enables
the first experimental calibration of the temperature scale used
in experiments on fermionic pair condensation, where the ideal
Fermi gas temperature is measured before sweeping the magnetic
field to the strongly interacting regime. Our calibration shows
that the ideal gas temperature measured for the onset of pair
condensation corresponds closely to  the critical temperature $T_c$ estimated in the
strongly interacting regime from the fits to our $E(S)$ data. We
also calibrate the empirical temperature employed in studies of
the heat capacity and obtain nearly the same $T_c$. We determine
the ground state energy by three different methods, using sound
velocity measurements, by extrapolating $E(S)$ to $S=0$ and by
measuring the ratio of the cloud sizes in the strongly and weakly
interacting regimes. The results are in very good agreement with
recent predictions. Finally, using universal thermodynamic
relations, we estimate the chemical potential and heat capacity of
the trapped gas from the $E(S)$ data.

\end{abstract}

\section{Introduction}

Interacting fermionic particles play a central role in the
structure of matter and exist over a very broad range of energies,
from extremely low temperature trapped atomic Fermi gases, where
$T<10^{-7}$ K~\cite{ohara02,thomas04}, to very high temperature
primordial matter, like quark-gluon plasmas, where $T>10^{12}$
K~\cite{heinz03}. For all of these systems, the most intriguing
physics is related to very strong interactions between fermionic
particles, such as the strong coupling between electrons in
high-$T_c$ superconductors and the strong interactions between
neutrons in neutron matter.

Current many-body quantum theories face great challenges in
solving problems for strongly interacting Fermi systems, due to
the lack of a small coupling parameter. For example, the critical
temperature of a superfluid-normal fluid transition in a strongly
interacting Fermi gas has been controversial for many years. The
critical temperature $T_c/T_F$ has been predicted to have values
in the range between 0.15 and 0.35 by different theoretical
methods~\cite{DrummondUniversal,DrummondComparative,ZwergerUnitaryGas,BulgacThermodynamics,chen05thermodynamics,Torma,
BruunViscous,perali04}.  A complete understanding of the physics
of strongly interacting systems can not yet be obtained from a
theoretical point of view. There is a pressing need to investigate
strongly interacting fermions experimentally.

In recent years, based on progress in optical cooling and trapping
of fermionic atoms, a clean and controllable strongly interacting
Fermi  system, comprising a degenerate, strongly interacting Fermi
gas~\cite{ohara02,thomas04}, is now of interest to the whole physics
community. Strongly interacting Fermi gases are produced near a
broad Feshbach resonance~\cite{ohara02,houbiers98,Luo06Cooling}, where
the zero energy s-wave scattering length $a_S$ is large compared
to the interparticle spacing, while the interparticle spacing is
large compared to the range of the two-body interaction. In this
regime, the system is known as a unitary Fermi gas, where the
properties are universal and independent of the details of the
two-body scattering interaction~\cite{heiselberg01,ho04}. In
contrast to other strongly interacting Fermi systems, in atomic
gases, the interactions, energy, and spin population can be
precisely adjusted, enabling a variety of experiments for
exploring this model system.

Intense studies of strongly interacting Fermi gases have been
implemented over the past several years from a variety of
perspectives. Some of the first experiments observed the expansion
hydrodynamics of the strongly interacting
cloud~\cite{ohara02,bourdel03}. Evidence for superfluid
hydrodynamics was first  observed in collective
modes~\cite{kinast04a,bartenstein04}. Collective modes were later
used to study the $T=0$ equation of state throughout the crossover
regime~\cite{kinast05b,Altmeyer07,Wright07}. Recently,
measurements of sound velocity have also been used to explore the
$T=0$ equation of state~\cite{Joseph07Sound}.  Below a Feshbach
resonance, fermionic atoms join to form stable molecules and
molecular Bose-Einstein
condensates~\cite{greiner03,jochim03,zwierlein03,regal04,zwierlein04}.
Fermionic pair condensation has been observed by projection
experiments using fast magnetic field
sweeps~\cite{regal04,zwierlein04}. Above resonance, strongly bound
pairs have been probed by radio frequency and optical
spectroscopy~\cite{chin04,partridge05,schunck08,JinPhotoemission}.
Phase separation has been observed in spin polarized
samples~\cite{zwierlein05,partridge05a}. Rotating Fermi gases have
revealed vortex lattices in the superfluid
regime~\cite{zwierlein05b,Schunck07} as well as irrotational flow
in both the superfluid and normal fluid regimes~\cite{Clancy07}.
Measurement of the thermodynamic properties of a strongly
interacting Fermi gas was first accomplished by  adding a known
energy to the gas, and then determining an empirical temperature
that was calibrated using a pseudogap theory~\cite{kinast05a}.
Recent model-independent measurements of the energy and
entropy~\cite{Luo07Entropy} provide a very important piece of the
puzzle, because they enable direct and precision tests that
distinguish predictions from recent many-body theories, without
invoking any specific theoretical
model~\cite{DrummondUniversal,DrummondComparative}.

One of the major challenges for the experiments in strongly
interacting Fermi gases is the lack of a precise model-independent
thermometry. Two widely-used  thermometry methods are
model-dependent, in that they rely on theoretical models for
calibration. The first relies on adiabatic magnetic field sweeps
between the molecular BEC regime and the strongly interacting
regime~\cite{chin04,chen06phase}. Subsequently, the temperature of
the strongly interacting gas is estimated from the measured
temperature in the BEC  regime using a theoretical model of the
entropy~\cite{chen05thermodynamics}. The second method, used by
our group~\cite{kinast05a}, is based on determining an empirical
temperature from the cloud profiles that is calibrated by comparing the measured density
distribution with a theoretical model for the density profiles.
Currently two model-independent thermometry methods have been
reported for strongly-interacting gases. One is the technique
employed by the MIT group~\cite{zwierlein06}, which is only
applicable to imbalanced mixtures of spin-up and spin-down atoms.
That method is based on fitting the noninteracting edge for the
majority spin after phase separation. Another model-independent
method is demonstrated in Ref.~\cite{Luo07Entropy}, which is
applicable to both  balanced and imbalanced mixtures of spin-up
and spin-down fermions. The energy $E$ and entropy $S$ are
measured and then parameterized to determine a smooth curve
$E(S)$.  Then the temperature in both the superfluid and normal
fluid regime is obtained from the fundamental thermodynamic
relation $T=\partial E/\partial S$.

In this paper, we will describe our model-independent
thermodynamic experiments on a strongly interacting Fermi gas of
$^6$Li, which we have conducted at Duke University. First, we will
describe our measurements of both the total energy $E$ and the
total entropy $S$  of a trapped strongly-interacting Fermi gas
tuned near a Feshbach resonance. Then, we determine the
temperature $T=\partial E/\partial S$ after showing that the
$E(S)$ data are very well parameterized by using two different
power laws that are joined with continuous $E$ and $T$ at a
certain entropy $S_c$ that gives the best fit. To examine the
sensitivity of the temperature to the form of the fit function, we
employ two different fit functions that allow for a heat capacity
jump or for a continuous heat capacity at $S_c$. We find that the
$T$  values closely agree for both cases. We find a significant
change in the scaling of $E$ with $S$ above and below $S_c$, in
contrast to the behavior for an ideal Fermi gas, where a single
power-law well parameterizes $E(S)$ over the same energy range. By
interpreting $S_c$ as the critical entropy for a superfluid-normal
fluid transition in the strongly interacting Fermi gas, we
estimate the critical energy $E_c$ and critical temperature $T_c$.
Both the model-independent $E(S)$ data and the estimated critical
parameters are compared with several recent many-body theories
based on both analytic and quantum Monte Carlo methods.

We also show how parameterizing the $E(S)$ data provides
experimental temperature calibrations, which helps to unify, in a
model-independent way, the results obtained by several
groups~\cite{regal04,zwierlein04,JinPhotoemission,kinast05a,Luo07Entropy}.
First we relate the endpoint temperatures for adiabatic sweeps of
the bias magnetic field between the strongly interacting and ideal
noninteracting regimes, as used in the JILA experiments to
characterize the condensed pair fraction~\cite{regal04,JinPhotoemission}.
This enables the ideal gas temperature observed for the onset of
pair condensation~\cite{regal04,JinPhotoemission} to be related to
the critical temperature of the strongly interacting Fermi gas.
The temperature obtained by parameterizing the strongly
interacting gas data also calibrates the empirical temperature
based on the cloud profiles, as used in our previous studies of
the heat capacity~\cite{kinast05a}. These temperature calibrations
yield values of $T_c$ close to that estimated from our $E(S)$ data.

Next, we discuss three different methods for determining the
universal many-body parameter, $\beta$~\cite{ohara02}, where
$1+\beta$ is the energy per particle in a uniform strongly
interacting Fermi gas at $T=0$ in units of the energy per particle
of an ideal Fermi gas at the same density. First, we describe the
measurement of the sound velocity at resonance and its
relationship to $\beta$. Then, we determine $\beta$ from the
ground state energy $E_0$ of the trapped gas. Here, $E_0$ is
obtained by extrapolating the $E(S)$ data to $S=0$, as suggested
by Hu et al.~\cite{DrummondUniversal}. This avoids a systematic
error in the sound velocity experiments arising from the unknown
finite temperature. Finally, to explore the systematic error
arising from the measurement of the number of atoms, $\beta$ is
determined in a number-independent manner from the ratio of the
cloud sizes in the strongly and weakly interacting regimes. All
three results are found to be in very good agreement with each
other and with recent predictions.

Finally, we obtain three universal thermodynamic functions from
the parameterized $E(S)$ data, the energy  $E(T)$, heat capacity
$C(T)$, and global chemical potential $\mu_g(E)$.

\section{Experimental Methods}

Our experiments begin with an optically-trapped highly degenerate,
strongly interacting Fermi gas of $^6$Li~\cite{ohara02}. A 50:50
mixture of the two lowest hyperfine states of $^6$Li atoms is
confined in an ultrastable CO$_2$ laser trap with a bias magnetic
field of 840 G, just above a broad Feshbach resonance at $B=834$
G~\cite{bartenstein05}. At 840 G, the gas is cooled close to the
ground state by lowering the trap depth
$U$~\cite{ohara02,Luo06Cooling}. Then $U$ is recompressed to a
final trap depth of $U_0/k_B=10\,\mu$K, which is much larger than
the energy per particle of the gas, for the highest energies
employed in the experiments. This suppresses evaporation during
the time scale of the measurements. The shallow trap yields a
low density that suppresses three body loss and heating. The low
density  also yields a weakly interacting sample when the bias
magnetic field is swept to 1200 G, although the scattering length
is $-2900$ bohr, as discussed in detail in \S~\ref{sec:entropy}.

The shape of the trapping potential is that of a gaussian laser
beam, with a transverse gaussian profile determined by the spot
size and an axial lorentzian profile determined by the Rayleigh
length. To simplify the calculations of the ideal gas properties
in subsequent sections, as well as the theoretical modelling, we
take the trap potential to be approximated by a three dimensional
gaussian profile,
\begin{equation}
U(x,y,z)=U_0\,\left(1-\exp\left(-\frac{2x^2}{a_x^2}-\frac{2y^2}{a_y^2}-\frac{2z^2}{a_z^2}\right)\right),
\label{eq:Udipole}
\end{equation}
where $a_{x,y,z}$ is the $1/e^2$ width of trap for each direction.
Here, we take the zero of energy to be at $\mathbf{r}=0$. When the
cold atoms stay in the deepest portion of the optical trap, where
$x(y,z)<<a_x(a_y,a_z)$, the gaussian potential can be well
approximated as a harmonic trap with transverse frequencies
$\omega_x$, $\omega_y$ and axial frequency $\omega_z$, where
\begin{equation}
\label{eq:TrapFreqWellDepth}
\omega_{x,y,z}=\sqrt{\frac{4\,U_0}{m\, a_{x,y,z}^2}}.
\end{equation}
Here $m$ is the $^6$Li mass. At our final trap depth $U_0$, the
measured transverse frequencies  are $\omega_x=2\pi\times 665(2)$
Hz and $\omega_y=2\pi\times 764(2)$ Hz. The axial frequency is
weakly magnetic field dependent since the total axial frequency
has both an optical potential contribution $\omega^2_{oz}$
determined by Eq.~\ref{eq:TrapFreqWellDepth} and a magnetic
potential contribution arising from magnetic field curvature,
$\omega^2_{mz}$. The net axial frequency is then
$\omega_z=\sqrt{\omega_{oz}^2+\omega_{mz}^2}$. We find
$\omega_z=2\pi\times 30.1(0.1)$ Hz at 840 G and
$\omega_z=2\pi\times 33.2(0.1)$ Hz at 1200 G. The total number of
atoms is $N\simeq 1.3\times 10^5$. The corresponding Fermi energy
$E_F$ and Fermi temperature $T_F$ at the trap center for an ideal
noninteracting harmonically trapped gas are $E_F=k_B
T_F\equiv\hbar\,\bar{\omega}(3N)^{1/3}$, where
$\bar{\omega}=(\omega_x\omega_y\omega_z)^{1/3}$. For our trap
conditions, we obtain $T_F\simeq 1.0\,\mu$K.

Using $\bar{\omega}$, we can rewrite Eq.~\ref{eq:Udipole} as a
symmetric effective potential,
\begin{equation}
\label{eq:reducedTruePotential}
U(\mathbf{r})=U_0\,\left(1-\exp\left(-\frac{m \bar{\omega}^2
\mathbf{r}^2 }{2U_0}\right)\right),
\end{equation}
where $\mathbf{r}$ is the scaled position vector. Here,
$\mathbf{r}^2=\tilde{x}^2+\tilde{y}^2+\tilde{z}^2$ with
$\tilde{x}=\omega_xx/\bar{\omega},\tilde{y}=\omega_yy/\bar{\omega},\tilde{z}=\omega_zz/\bar{\omega}$.
To obtain the anharmonic corrections for the gaussian trap, we
expand Eq.~\ref{eq:reducedTruePotential} in a Taylor series up to second order
in $\mathbf{r}^2$,
\begin{equation}
\label{eq:TruePotentialQuartic} U(\mathbf{r})=\frac{m
\bar{\omega}^2 \mathbf{r}^2 }{2}-\frac{m^2 \bar{\omega}^4
\mathbf{r}^4}{8U_0}.
\end{equation}

\subsection{Energy Measurement}
Model-independent energy measurement is based on a virial theorem,
which for an ideal gas in a harmonic confining potential $U_{ho}$
yields $E=2\langle U_{ho}\rangle$. Since the harmonic potential
energy is proportional to the mean square cloud size, measurement
of the cloud profile determines the total energy. Remarkably, a
trapped unitary Fermi gas at a broad Feshbach resonance obeys the
same virial theorem as an ideal gas, although it contains
superfluid pairs, noncondensed pairs, and unpaired atoms, all
strongly interacting. This has been demonstrated both
theoretically and experimentally~\cite{thomas05}. The virial
theorem shows that the total energy of the gas at all temperatures
can be measured from the cloud profile using
\begin{equation}
E=\left\langle U+\frac{1}{2}\,\mathbf{x}\cdot\nabla
U\right\rangle, \label{eq:energymeas}
\end{equation}
where $U$ is the trapping potential and $\mathbf{x}$ is the
position vector.  Eq.~\ref{eq:energymeas} can be shown to be valid
for any trapping potential $U$ and for any spin mixture, without
assuming either the local density approximation or harmonic
confinement~\cite{CastinVirial,Son07,WernerVirial,ThomasVirial08}.

Using Eq.~\ref{eq:TruePotentialQuartic} in Eq.~\ref{eq:energymeas}
and keeping the lowest order  anharmonic corrections, we obtain
the energy per atom in terms of the axial mean square size,
\begin{equation}
E=3m\omega_z^2\left\langle
z^2\right\rangle\left[1-\frac{5}{8}\frac{m\omega_z^2\left\langle
z^4\right\rangle}{U_0\langle z^2\rangle}\right].
\label{eq:energygaussian}
\end{equation}
Here,  we have used the local density approximation with a scalar
pressure, which ensures that $\langle x\partial U/\partial
x\rangle=\langle y\partial U/\partial y\rangle=\langle z\partial
U/\partial z\rangle$. For the ground state, where the spatial
profile is a zero temperature Thomas-Fermi profile, we have
$\langle z^4\rangle = 12\langle z^2\rangle^2/5$. For energies
$E/E_F >1$, where the spatial profile is approximately gaussian,
we have $\langle z^4\rangle = 3\langle z^2\rangle^2$. Since the
anharmonic correction is small at low temperatures where the cloud
size is small, we use the gaussian approximation over
the whole range of energies explored in our experiments.  For the
conditions of our experiments, there is no evidence that the local
density approximation breaks down for a 50:50 spin mixture. In
this case, measurement of the mean square size in any one
direction determines the total energy. From
Eq.~\ref{eq:energygaussian}, we see that by simply measuring the
axial mean square size $\langle z^2 \rangle_{840}$ at 840 G and
measuring the axial trap frequency by parametric resonance, we
actually measure $E_{840}$, the total energy per particle of the
strongly interacting Fermi gas at 840 G. This determines the total
energy per particle in a model-independent
way~\cite{Luo07Entropy}.

\subsection{Entropy Measurement}
\label{sec:entropy}

The entropy $S$ of the strongly interacting gas at 840 G is
determined by adiabatically  sweeping the bias magnetic field from
840 G to 1200 G, where the gas is weakly
interacting~\cite{Luo07Entropy}. The entropy $S_W$ of the weakly
interacting gas is essentially the entropy of an ideal Fermi gas
in a harmonic trap, which can be calculated in terms of the mean
square axial cloud size $\langle z^2\rangle_{1200}$ measured after
the sweep. Since the sweep is adiabatic, we have
\begin{equation}
S=S_W.
\label{eq:adiabatic}
\end{equation}

The adiabaticity of the magnetic field sweep is verified by
employing a round-trip-sweep: The mean square size of the cloud at
840 G after a round-trip-sweep lasting 2s is found to be within
3\%  of mean square size of a cloud that remains at 840 G for a
hold time of 2s. The nearly unchanged atom number and mean square
size proves the sweep does not cause any significant atom loss or
heating, which ensures entropy conservation for the sweep. The
background heating rate is the same with and without the sweep and
increases the mean square size by about 2\% over 2s. The mean
square size data are corrected by subtracting the increase arising
from background heating over the 1 s sweep
time~\cite{Luo07Entropy}.

At 1200 G in our shallow trap, we have $k_F\,a_S=-0.75$, where the
Fermi wavevector $k_F=(2mE_F/\hbar^2)^{1/2}$ and the s-wave
scattering length $a_S = -2900$ bohr~\cite{bartenstein05}. We find
that the gas is weakly interacting: For the lowest temperatures
attained in our experiments, the gas at 1200 G is a normal fluid
that we observe to expand ballistically.  We have calculated
the ground state mean square size at 1200 G in our gaussian trap,
based on a mean-field theory, $\langle
z^2\rangle_{W0}=0.69\,z_{F}^2(1200G)$~\cite{Luo07Entropy}, which
is close to that of an ideal harmonically trapped gas, $\langle
z^2\rangle_{I0}=0.75\,z_F^2(1200G)$. Here, $z_F^2(B)$ is the mean
square size corresponding to the Fermi energy of an ideal
noninteracting Fermi gas at magnetic field $B$, which includes the
magnetic field dependence of the axial trapping frequency:
$E_F(B)\equiv 3m\omega_z^2(B)z_F^2(B)$.

We expect that the entropy
of the gas at 1200 G is close to that of an ideal gas, except for
a mean field shift of the energy. We therefore assume that a reasonable
approximation to the entropy is that of an ideal Fermi gas,
$S_I(\langle z^2\rangle_I-\langle
z^2\rangle_{I0})$, where $\langle z^2\rangle_{I0}$ is the  ground
state mean square size of an ideal Fermi gas in the gaussian
trapping potential of Eq.~\ref{eq:reducedTruePotential}. Here, we
apply an elementary calculation based on integrating the density
of states for the gaussian trap with the entropy per orbital $s
=-k_B[f\ln f+(1-f)\ln(1-f)]$, where  $f(\epsilon)$ is the ideal
Fermi gas occupation number at temperature $T$ for an orbital of
energy $\epsilon$. By calculating $S_I$ as a function of the {\it
difference} between the finite temperature and ground state mean
square cloud sizes, we reduce the error arising from the mean
field shift at 1200 G, and ensure that $S_I=0$ for the ground
state.

The exact entropy of a weakly interacting gas $S_W$ at 1200G,
$S_{W}(\langle z^2\rangle_{W}-\langle z^2\rangle_{W0})$, has been
calculated using many-body theories~\cite{DrummondUniversal,BulgacThermodynamics} for
the gaussian potential of Eq.~\ref{eq:reducedTruePotential}. In
the experiments, we determine the value of $\langle
z^2\rangle_W-\langle z^2\rangle_{W0}$, where we take $\langle
z^2\rangle_{W0}=0.71\,z_F^2(1200)$, the value measured at our
lowest energy at 1200 G by extrapolation to $T=0$ using the
Sommerfeld expansion for the spatial profile of an ideal gas. This
result is close to the theoretical value, $0.69\,z_F^2(1200)$.

\begin{figure}
\centerline{\includegraphics[width=
4.0in,clip]{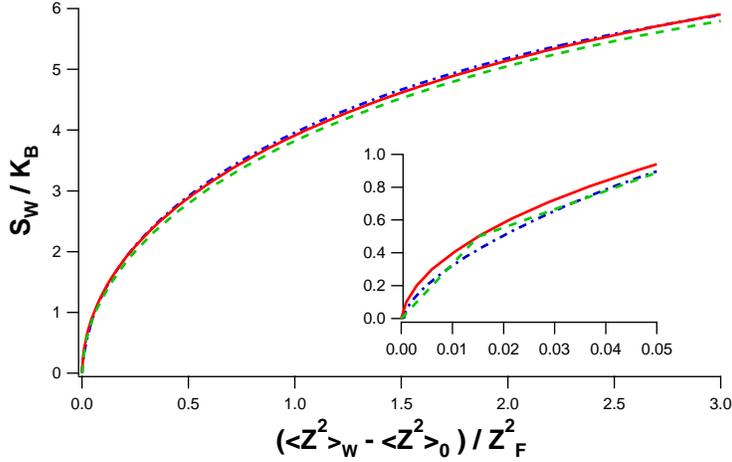}} \caption {Color online. Comparison of the
entropy versus mean square size curves of a weakly interacting
Fermi gas with $k_F a_S=-0.75$ in a gaussian trap with that of a
noninteracting Fermi gas. Many-body predictions are shown as the green dashed curve~\cite{DrummondUniversal} and the blue dot-dashed curve~\cite{BulgacThermodynamics}. The red solid curve is our
calculation for a noninteracting gas~\cite{luothesis}. All of the calculations
employ the gaussian potential  of
Eq.~\ref{eq:reducedTruePotential} with a trap depth $U_0/E_F=10$.
The points where $S=0$ automatically coincide, since the entropy
is determined in all cases as a function of mean square size
relative to that of the ground state. This corrects approximately
for the mean-field shift of the mean square size. Note that for a
weakly interacting gas, $\langle z^2\rangle_0$ is the calculated
ground cloud size for each theory that includes the mean field
energy shift, while for the noninteracting case $\langle
z^2\rangle_0$ is  $\langle z^2\rangle_{I0}$, the value for an
ideal trapped Fermi gas. $z_F^2$ is the mean square size for an
energy $E=E_F$.\label{fig:entropysizeoverlap}}
\end{figure}

The entropy versus cloud size curve for an ideal noninteracting
Fermi gas and the exact value for a weakly interacting gas $S_W$
at 1200 G are plotted in Fig.~\ref{fig:entropysizeoverlap}. We
find that the entropies $S_{W}(\Delta \langle z^2\rangle =\langle
z^2\rangle_{W}-\langle z^2\rangle_{W0})$ and $S_I(\Delta \langle
z^2\rangle)$),  agree within a few percent
 over most of the energy range we studied, except at the point
of lowest measured energy, where they differ by 10\%. The results
show clearly that the shape of the entropy curve of a weakly
interacting Fermi gas is nearly identical to that of an ideal gas
when the mean field shift of the ground state size is included by
referring the mean square cloud size to that of the ground state.
So we have to a good approximation,
\begin{equation}
S = S_{W}(\langle z^2\rangle_{W}-\langle z^2\rangle_{W0})
\simeq S_I(\langle z^2\rangle_{W}-\langle z^2\rangle_{W0}).
\end{equation}
Since the corrections to ideal gas behavior are small, the
determination of $S_{1200}$ by measuring the axial mean square
size $\langle z^2\rangle_{1200}$ relative to the ground state
provides an essentially model-independent estimate of the entropy
of the strongly interacting gas.

\subsection{Sound Velocity Measurement}
Sound velocity measurements have been implemented for Fermi gases
that are nearly in the ground state, from the molecular BEC regime
to the weakly interacting Fermi gas regime~\cite{Joseph07Sound}. A
sound wave is excited in the sample by using a thin slice of green
light that bisects the cigar-shaped cloud. The green light at 532
nm is blue detuned from the 671 nm transition in lithium, creating
a knife that locally repels the atoms. The laser knife is pulsed
on for 280 $\mu$s, much shorter than typical sound propagation
times $\sim 10$ ms and excites a ripple in the density consisting
of low density valleys and high density peaks. After excitation,
the density ripple propagates outward along the axial direction
$z$. After a variable amount of propagation time, we release the
cloud, let it expand, and image destructively. In the strongly
interacting regime, we use zero-temperature Thomas-Fermi profiles
for a non-interacting Fermi gas to fit the density profiles, and
locate the positions of the density valley and peak. By recording
the position of the density ripple versus the propagation time,
the sound velocity is determined. A detailed discussion of
potential sources of systematic error is given by Joseph et
al.~\cite{Joseph07Sound}.

For a strongly interacting Fermi gas in the unitary limit, the
sound velocity $c_0$ at the trap center   for the ground state is
determined by the Fermi velocity of an ideal gas at the trap
center, $\textrm{v}_F=(2E_F/m)^{1/2}$ and the universal constant
$\beta$,
\begin{equation}
\frac{c_0}{\textrm{v}_F}=\frac{(1+\beta)^{1/4}}{\sqrt{5}}.
\label{eq:soundres}
\end{equation}
A precision measurement of the sound speed therefore enables a
determination of $\beta$~\cite{Joseph07Sound}. As discussed below,
the values of $\beta$ determined from the $E(S)$ data and the
sound velocity data are in very good agreement.

\section{Comparison of Thermodynamic Data with Theory}
In the experiments, the raw data consists of the measured mean
square cloud sizes at 840 G and after an adiabatic sweep of the
magnetic field to 1200 G. Using this data, we determine both the
energy and the entropy of the strongly interacting gas. The data
is then compared to several recent predictions.

\subsection{Mean Square Cloud Sizes}
We begin by determining the axial mean square cloud sizes at 840 G
and after the adiabatic sweep to 1200 G. Since the atom number can
vary between different runs  by up to 20\%, it is important to
make the comparison independent of the atom number and trap
parameters. For this purpose, the mean square sizes are given in
units of  $z_F^2(B)$, as defined above.

The measured mean square sizes are listed in
Table~\ref{tbl:energyentropy}.  In the experiments, evaporative
cooling is used to produce an atom cloud near the ground state.
Energy is controllably added by releasing the cloud and then
recapturing it after a short time $t_{heat}$ as described
previously~\cite{Luo07Entropy}.  For a series of different values
of  $t_{heat}$, the energy at 840 G is directly measured from the
axial cloud size according to Eq.~\ref{eq:energygaussian}. Then
the same sequence is repeated, but the cloud size is measured
after an adiabatic sweep to 1200 G. In each case, the systematic
increase in mean square size arising from background heating rate
is determined and subtracted. The total data comprise about 900
individual measurements of the cloud size at 840 G and 900 similar
measurements of the cloud size after a sweep to 1200 G. To
estimate the measurement error, we split the energy scale at 840 G
into bins with a width of $\Delta E=0.04\,E_F$. Measured data
points within the width of the energy bin are used to calculate
the average measured values of the cloud sizes and the
corresponding standard deviation at both 840 G and 1200 G.

\begin{table}
\begin{center}
\begin{tabular}{|l|c|c|c|c|c|c|}
\hline
\rule[-3pt]{0pt}{14pt} &$\langle z^2\rangle_{840}/z_F^2(840)$&$\langle z^2\rangle_{1200}/z_F^2(1200)$&$E_{840}/E_F$&$S_{1200}/k_B$&$S_{1200}^*/k_B$&$S_{1200}^{**}/k_B$\\
\hline \hline
\rule[-1pt]{0pt}{14pt}1&0.568(4)&0.743(6)&0.548(4)&0.63(8)&0.91(23)&0.97(5)\\
\hline
\rule[-1pt]{0pt}{14pt}2&0.612(5)&0.776(13)&0.589(5)&0.99(11)&1.18(22)&1.24(9)\\
\hline
\rule[-1pt]{0pt}{14pt}3&0.661(5)&0.803(11)&0.634(5)&1.22(8)&1.36(20)&1.42(7)\\
\hline
\rule[-1pt]{0pt}{14pt}4&0.697(9)&0.814(15)&0.667(8)&1.30(10)&1.43(18)&1.49(8)\\
\hline
\rule[-1pt]{0pt}{14pt}5&0.74(1)&0.87(4)&0.71(1)&1.6(2)&1.72(18)&1.8(2)\\
\hline
\rule[-1pt]{0pt}{14pt}6&0.79(1)&0.89(2)&0.75(1)&1.7(1)&1.79(15)&1.9(1)\\
\hline
\rule[-1pt]{0pt}{14pt}7&0.83(1)&0.94(2)&0.79(1)&2.0(1)&2.03(16)&2.1(1)\\
\hline
\rule[-1pt]{0pt}{14pt}8&0.89(2)&1.02(2)&0.84(2)&2.3(1)&2.32(18)&2.4(1)\\
\hline
\rule[-1pt]{0pt}{14pt}9&0.91(1)&1.02(3)&0.86(1)&2.3(1)&2.31(16)&2.4(1)\\
\hline
\rule[-1pt]{0pt}{14pt}10&0.97(1)&1.10(1)&0.91(1)&2.55(4)&2.57(17)&2.64(4)\\
\hline
\rule[-1pt]{0pt}{14pt}11&1.01(1)&1.17(2)&0.94(1)&2.74(7)&2.75(19)&2.82(7)\\
\hline
\rule[-1pt]{0pt}{14pt}12&1.05(1)&1.18(1)&0.98(1)&2.78(4)&2.80(17)&2.87(4)\\
\hline
\rule[-1pt]{0pt}{14pt}13&1.10(1)&1.22(1)&1.03(1)&2.89(2)&2.90(15)&2.97(2)\\
\hline
\rule[-1pt]{0pt}{14pt}14&1.25(2)&1.35(5)&1.15(1)&3.21(12)&3.20(14)&3.27(12)\\
\hline
\rule[-1pt]{0pt}{14pt}15&1.28(1)&1.39(3)&1.18(1)&3.28(6)&3.28(14)&3.35(6)\\
\hline
\rule[-1pt]{0pt}{14pt}16&1.44(2)&1.49(2)&1.31(2)&3.49(4)&3.48(9)&3.55(4)\\
\hline
\rule[-1pt]{0pt}{14pt}17&1.53(2)&1.62(6)&1.39(2)&3.74(10)&3.73(11)&3.80(10)\\
\hline
\rule[-1pt]{0pt}{14pt}18&1.58(1)&1.63(1)&1.42(1)&3.76(2)&3.74(8)&3.81(2)\\
\hline
\rule[-1pt]{0pt}{14pt}19&1.70(2)&1.73(6)&1.52(1)&3.94(9)&3.92(7)&3.99(9)\\
\hline
\rule[-1pt]{0pt}{14pt}20&1.83(5)&1.79(2)&1.62(4)&4.03(3)&4.01(1)&4.08(3)\\
\hline
\rule[-1pt]{0pt}{14pt}21&1.93(3)&1.96(3)&1.70(2)&4.28(4)&4.26(6)&4.32(4)\\
\hline
\rule[-1pt]{0pt}{14pt}22&2.11(5)&2.17(3)&1.83(4)&4.55(3)&4.53(7)&4.59(3)\\
\hline
\end{tabular}
\end{center}
\caption{Mean square axial cloud size, energy, and entropy
measured in a trapped strongly interacting Fermi gas. $\langle
z^2\rangle_{840}/z_F^2(840)$ is the scaled axial mean square size
at 840 G. $\langle z^2\rangle_{1200}/z_F^2(1200)$ is the scaled
axial mean square size at 1200 G. $E_{840}/E_F$ is the energy per
particle of a strongly interacting Fermi gas at 840 G, calculated
using Eq.~\ref{eq:energygaussian}. $S_{1200}/k_B$ is the
corresponding entropy per particle of the gas after an adiabatic
sweep of magnetic field from 840 G to 1200 G, where the
noninteracting entropy curve (for the gaussian trap) is used to
determine the entropy at 1200 G and the ground state mean square
size is assumed to be $\langle z^2\rangle_{0}=0.71\,z_F^2(1200)$.
$S_{1200}^{**}/k_B$ is the ideal gas entropy result assuming
$\langle z^2\rangle_{0}=0.69\,z_F^2(1200)$. $S_{1200}^*/k_B$ is
the entropy obtained using an exact many-body calculation for
$k_Fa=-0.75$~\cite{DrummondUniversal}.} \label{tbl:energyentropy}
\end{table}

\begin{figure}
\centerline{\includegraphics[width= 4.0in,clip]{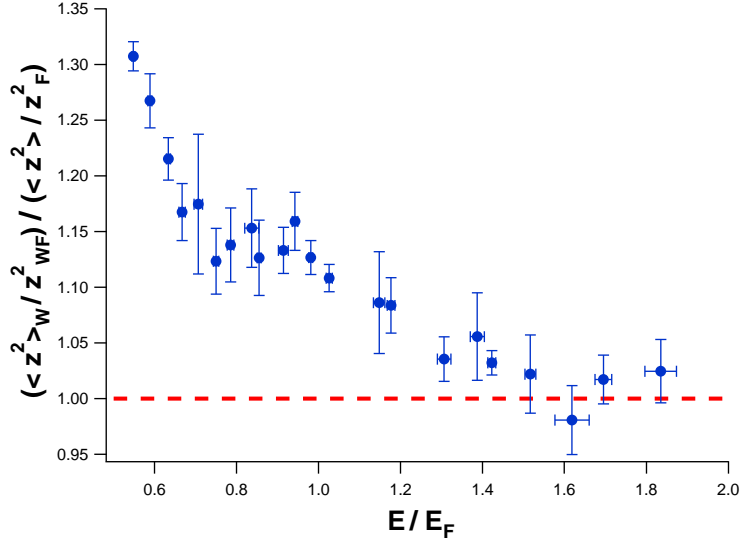}}
\caption {Color online. The ratio of the mean square cloud size at
1200 G, $\langle z^2\rangle_{1200}$, to that at 840 G, $\langle
z^2\rangle_{840}$, for an isentropic magnetic field sweep. $E$ is
the total energy per particle of the strongly interacting gas at
840 G and  $E_F$ is the ideal gas Fermi energy at 840 G. The ratio
converges to unity at high energy as expected (the dashed line).
Here, $z_{WF}$ is evaluated at 1200 G, while $z_F$ is evaluated at
840 G (see text).\label{fig:sizeratio}}
\end{figure}

The ratio of the mean square axial cloud size at 1200 G (measured
after the sweep) to that at 840 G (measured prior to the sweep) is
plotted in Fig.~\ref{fig:sizeratio} as a function of the energy of
a strongly interacting gas at 840 G. The ratio is $\geq 1$, since
for an adiabatic sweep of the magnetic field from the strongly
interacting regime to the weakly interacting regime, the total
entropy in the system is conserved but the energy increases: The
strongly interacting gas is more attractive than the weakly
interacting gas. A similar method was used to measure the
potential energy change in a Fermi gas of $^{40}$K, where the bias
magnetic field was adiabatically swept between the strongly
interacting regime at the Feshbach resonance and a noninteracting
regime above resonance~\cite{JinPotential}. The resulting
potential energy ratios are given as a function of the temperature
of the noninteracting gas~\cite{JinPotential}. In contrast, by
exploiting the virial theorem which holds for the unitary gas, we
determine both the energy and entropy of the strongly interacting
gas, as described below.

\subsection{Energy versus Entropy}

\begin{figure}
\centerline{\includegraphics[width=
4.0in,clip]{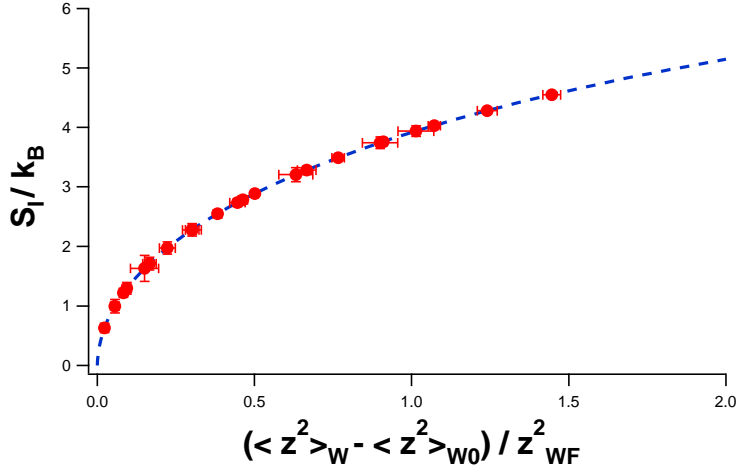}} \caption {Color online. The
conversion of the mean square size at 1200 G to the entropy. The
dashed line is the calculated entropy for a noninteracting Fermi
gas in the Gaussian trap with $U_0/E_F=10$. $\langle
z^2\rangle_0=0.71\,z_F^2$ is the measured ground state size for a
weakly interacting Fermi gas. The calculated error bars of the
entropy are determined from the measured error bars of the cloud
size at 1200 G. Here, $z_{WF}$ is evaluated at 1200
G.\label{fig:sizeconvertentropy}}
\end{figure}

Fig.~\ref{fig:sizeconvertentropy} shows the entropy which is
obtained from the mean square size at 1200 G $\langle
z^2\rangle_{1200}/z_F^2(1200)$ as listed in
Table~\ref{tbl:energyentropy}. First, we find the mean square size
relative to that of the ground state,  $(\langle
z^2\rangle_{1200}-\langle z^2\rangle_0)/z_F^2(1200)$. We use the
measured value $\langle z^2\rangle_0=0.71\,z_F^2(1200)$ for the
lowest energy state that we obtained at 1200 G, as determined by
extrapolation to $T=0$ using a Sommerfeld expansion for the
spatial profile of an ideal Fermi gas. Then we determine the
entropy in the noninteracting ideal Fermi gas approximation:
$S_I[\langle z^2\rangle_{1200}-\langle z^2\rangle_0]$, where we
have replaced $\langle z^2\rangle_{I0}$ by the ground state value
at 1200 G. As discussed above, this method automatically ensures
that $S=0$ corresponds to the measured ground state $\langle
z^2\rangle_0$ at 1200 G, and compensates for the mean field shift
between the measured $\langle z^2\rangle_0$ for a weakly
interacting Fermi gas and that calculated $\langle z^2\rangle_{I0}
= 0.77\,z_F^2$ for an ideal Fermi gas in our gaussian trapping
potential. As shown in Fig.~\ref{fig:entropysizeoverlap}, the
entropy obtained from a more precise many-body calculations is in
close agreement with the ideal gas entropy calculated in the ideal
gas approximation. The energy is determined from the cloud
profiles at 840 G using Eq.~\ref{eq:energygaussian}.

\begin{figure}
\centerline{\includegraphics[width=4.0in,clip]{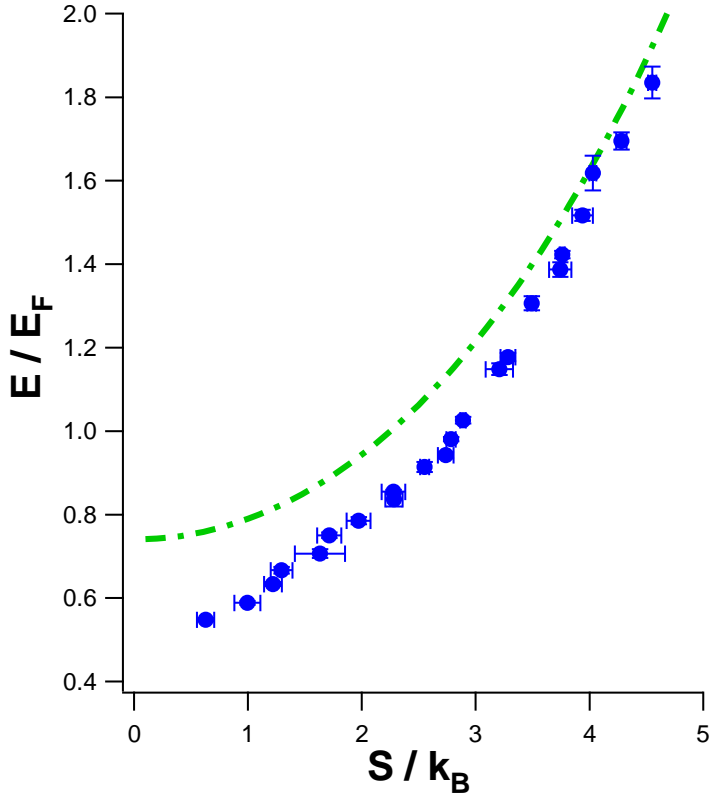}}
\caption {Color online. Measured total energy
per particle of a strongly interacting Fermi gas at 840 G versus
its entropy per particle.  For comparison,
the dot-dashed green curve shows $E(S)$ for an ideal Fermi gas.
For this figure, the ideal gas
approximation to the entropy is used, $S_{1200}/k_B$ of
Table~\ref{tbl:energyentropy}. \label{fig:energyentropy}}
\end{figure}

Finally, we generate the energy-entropy curve for a strongly
interacting Fermi gas, as shown in Fig.~\ref{fig:energyentropy}.
Here, the energy $E$ measured from the mean square axial cloud
size at 840 G is plotted as a function of the entropy $S$ measured
at 1200 G after an adiabatic sweep of the magnetic field. We note
that above $S=4\,k_B$ ($E=1.5\,E_F$) the $E(S)$ data (blue dots)
for the strongly interacting gas appear to merge smoothly to the
ideal gas curve (dashed green).

In addition to the entropy calculated in the ideal gas
approximation, Table~\ref{tbl:energyentropy} also provides a more
precise entropy $S^*(1200)$ versus the axial mean square cloud
size. These results are obtained by Hu et
al.~\cite{DrummondUniversal} using a many-body calculation for
$k_Fa=-0.75$ at 1200 G in the gaussian trap of
Eq.~\ref{eq:reducedTruePotential}.

\subsection{Testing Predictions from Many-Body Theories}

Perhaps the most important application of the energy-entropy
measurements is to test strong coupling many-body theories and
simulations. Since the energy and entropy are obtained in absolute
units without invoking any specific theoretical model, the data
can be used to distinguish recent predictions for a trapped
strongly interacting Fermi gas.

\begin{figure}[htb]
\centerline{\includegraphics[width=
3in,clip]{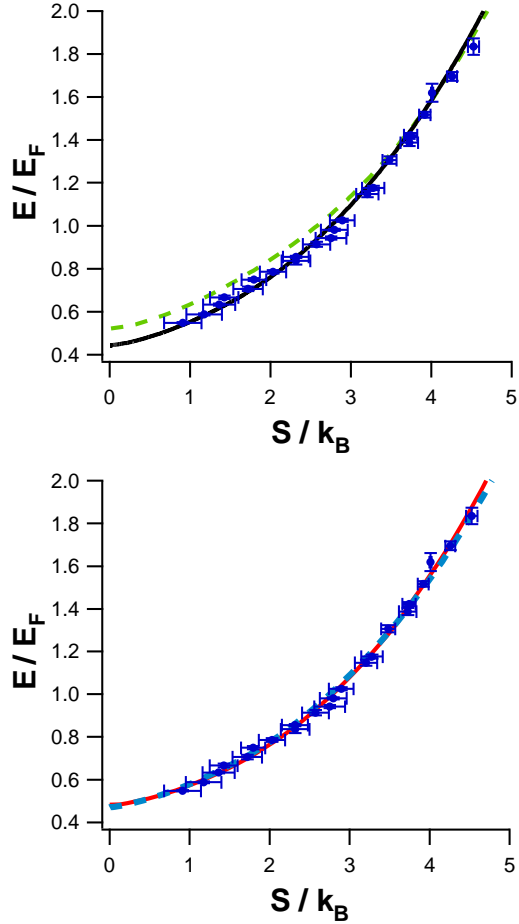}}\caption {Color online. Comparison of the
experimental energy versus entropy data with the calculations from strong
coupling many-body theories. Top: The green dashed curve is a pseudogap theory~\cite{chen05thermodynamics,QijinEntropy}.
The black solid curve is a LW-DDM variational calculation~\cite{ZwergerUnitaryGas}.
For this figure, the entropy is given by $S_{1200}^*/k_B$ in Table~\ref{tbl:energyentropy}.
Bottom: The blue dashed curve is an
NSR calculation~\cite{DrummondUniversal,DrummondComparative}. The
red solid curve is a quantum Monte Carlo
simulation~\cite{BulgacThermodynamics,BulgacUnitary}.
 For this figure, the entropy is given by $S_{1200}^*/k_B$ in Table~\ref{tbl:energyentropy}.
\label{fig:comparetheory}}
\end{figure}

Fig.~\ref{fig:comparetheory} shows how four different predictions
compare to the measured energy and entropy data. These include a
pseudogap theory~\cite{chen05thermodynamics,QijinEntropy}, a
combined Luttinger-Ward-De Dominicis-Martin (LW-DDM) variational
formalism~\cite{ZwergerUnitaryGas}, a T-matrix calculation using a
modified Nozi\`{e}res and Schmitt-Rink (NSR)
approximation~\cite{DrummondUniversal,DrummondComparative}, and  a
quantum Monte Carlo
simulation~\cite{BulgacThermodynamics,BulgacUnitary}. The most
significant deviations appear to occur near the ground state,
where the precise determination of the energy seems most
difficult. The pseudogap theory predicts a ground state energy
that is above the measured value while the prediction of
Ref.~\cite{ZwergerUnitaryGas} is somewhat low compared to the
measurement. All of the different theories appear to converge at
the higher energies.

\section{Determining the Temperature and Critical Parameters}
\label{sec:temp}

The temperature $T$ is determined from the measured $E(S)$ data
using the fundamental relation, $T=\partial E/\partial S$. To
implement this method, we need to parameterize the data to obtain
a smooth differentiable curve.

At low temperatures, one expects the energy to increase from the
ground state according to a power law in $T$ and a corresponding
power law in $S$, i.e.,  $E=E_0+a\,S^b$. For a harmonically
trapped ideal Fermi gas, we have in the Sommerfeld approximation
an energy per particle in units of $E_F$ given by
$E=3/4\,+\,\pi^2(T/T_F)^2/2$. The corresponding entropy per
particle in units of $k_B$ is $S=\pi^2(T/T_F)$, so that $E =
3/4\,+\,S^2/(2\pi^2)\simeq 0.75+0.05\,S^2$.

\subsection{Power Law Fit and Temperature for an Ideal Fermi Gas}
\label{sec:idealgas}

\begin{figure}
\centerline{\includegraphics[width=
4.0in,clip]{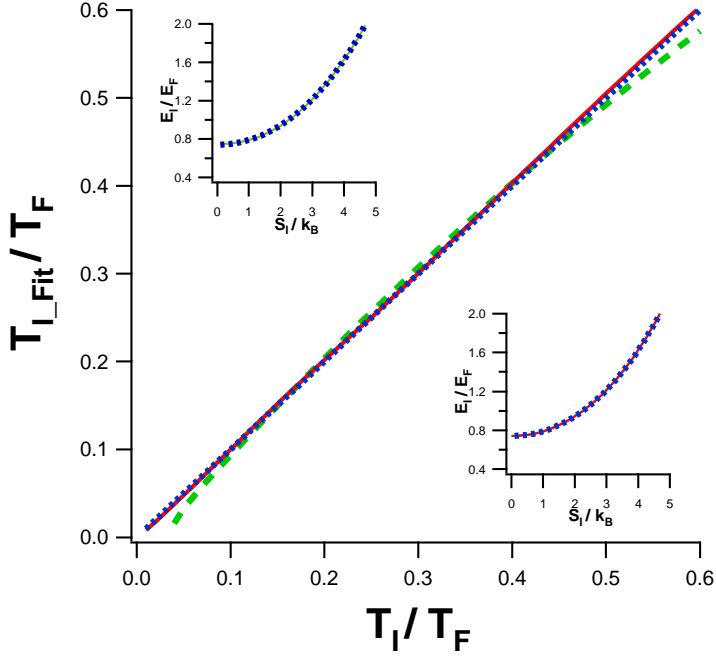}} \caption {Color online. The inset top left shows a single power-law fit
to the calculated energy versus entropy per particle of a
noninteracting Fermi gas in a gaussian trap. The exact result (blue
dots) is well fit by the single power law function (green solid
curve)
$E_I=(0.747\pm0.001)+(0.0419\pm0.0004)\,S^{(2.197\pm0.006)}$. The inset bottom right shows a two-power
law fit to the calculated energy versus entropy per particle of a
noninteracting Fermi gas in a gaussian trap. The exact
$E_I(S)$ (blue dots), is very well fit by the two power law
function (red solid curve) of Eq.~\ref{eq:EvsSlowengtwopower} for
$e\neq 0$ with $a=0.0480(2),b=2.08(3),S_c=2.16(3),d=2.483(7)$ and
$E_0=0.7415(2)$. The temperature of a noninteracting Fermi gas obtained from
the fits to the calculated energy versus entropy is compared to the
exact temperature.  The blue dotted line corresponds to exact
agreement. The green dot-dashed curve is from the single power law
fit showing deviation only at the highest and lowest temperatures.
The red curve is from two power law fit of
Eq.~\ref{eq:EvsSlowengtwopower} with $e\neq 0$ showing excellent
agreement over the full range.
\label{fig:IdealGasFit}}
\end{figure}

We attempt to use a single power law to fit the $E_I(S)$ curve for
a noninteracting Fermi gas in a gaussian trapping potential, with
$U_0/E_F=10$, as in our experiments. The energy and entropy are
calculated in the energy range $0.75\,E_F\leq E\leq 2\,E_F$ and
displayed as dots in Fig.~\ref{fig:IdealGasFit}. We find that
a single power law
$E_I=(0.747\pm0.001)+(0.0419\pm0.0004)\,S^{(2.197\pm0.006)}$ fits
the curve very well over this energy range. Note that the power
law exponent is $b\simeq 2$, close to the low temperature value.

Using the fit function, we can extract the reduced temperature
$T_{I,fit}/T_F=\partial E_I/\partial S_I$ as a function of $S_I$
and compare it to the theoretical reduced temperature $T/T_F$ at
the same $S_I$. The results are shown as the green dashed line in
Fig.~\ref{fig:IdealGasFit}. We see that the agreement is quite good
except below $T/T_F=0.1$ and above $0.5$, where the deviation is
$\simeq 10$\%.

To improve the fit and to make a more precise determination of the
temperature, we employ a fit function comprising two power laws
that are joined at a certain entropy $S_c$, which gives the best
fit. When used to fit the data for the strongly interacting Fermi
gas, we consider two types of fits that incorporate either a jump
in heat capacity or a continuous heat capacity at $S_c$. In this
way, we are able determine the sensitivity of the temperature and
critical parameters to the form of the fit function. The two types
of fits yield nearly identical temperatures, but different values
of $S_c$ and hence of the critical parameters, as discussed below.

We take the energy per particle $E$ in units of $E_F$ to be given
in terms of the entropy per particle in units of $k_B$ in the form
\begin{eqnarray}
E_<(S)&=&E_0+\,a\,S^b;\,\,\,\,0\leq S\leq S_c\nonumber\\
E_>(S)&=&E_1+\,c\,S^d+\,e(S-S_c)^2;\,\,\,\,S\geq S_c.
\label{eq:EvsSlowengtwopower0}
\end{eqnarray}

We constrain the values of $E_1$ and $c$ by demanding that energy
and temperature be continuous at the joining point $S_c$:
\begin{eqnarray}
E_<(S_c)&=&E_>(S_c)\nonumber\\
\left(\frac{\partial E_<}{\partial
S}\right)_{S_c}&=&\left(\frac{\partial E_>}{\partial
S}\right)_{S_c}. \label{eq:boundarycond}
\end{eqnarray}
By construction, the value of $e$ does not affect these constraints and is chosen in one of two ways. Fixing $e=0$, the fit incorporates a heat capacity jump at $S_c$, which arises from the change in the power law exponents at $S_c$. Alternatively, we choose $e$ so that the second derivative $E''(S)$ is continuous at $S_c$, making the heat capacity continuous.
The final fit function has 5 independent parameters $E_0,S_c,a,b,d$, and takes the form
\begin{eqnarray}
E_<(S)&=&E_0+\,a\,S^b;
\,\,0\leq S\leq S_c\nonumber\\
E_>(S)&=&E_0+\,a\,S_c^b\left[1-\frac{b}{d}+\frac{b}{d}\left(\frac{S}{S_c}\right)^d\right]+e(S-S_c)^2;
\,\,S\geq S_c.
 \label{eq:EvsSlowengtwopower}
\end{eqnarray}
Here, when $e$ is not constrained to be zero, it is given by
\begin{equation}
e = \frac{ab}{2}(b-d)S_c^{b-2}
\label{eq:e}
\end{equation}
Fig.~\ref{fig:IdealGasFit} shows the improved fit to the
calculated energy versus entropy of a noninteracting Fermi gas in
a gaussian trap for $U_0/E_F=10$, using
Eq.~\ref{eq:EvsSlowengtwopower} with $e\neq 0$, since the ideal
gas has no heat capacity jump. In this case, both power law
exponents $b$ and $d$ are close to $2$ as for the single power law
fit.  The temperature determined from the fit agrees very closely
with the exact temperature, as shown in Fig.~\ref{fig:IdealGasFit} (red
solid line).

\subsection{Power Law Fit and Temperature of a Strongly Interacting Fermi Gas}

In contrast to the noninteracting case, we have found that the
energy-entropy data of a strongly interacting Fermi gas is not
well fit by a single power law function~\cite{Luo07Entropy}.
However, the two power-law function fits quite well, with a factor
of two smaller value of $\chi^2$ than for the single power-law
fit. Here, we use $\chi^2=\sum_i (\frac{y-y_i}{\sigma_i})^2$,
where $y$~($y_i$) is the fitted~(data) value for the $i^{th}$
point, and $\sigma_i$ is the corresponding the standard error.

Motivated by the good fits of the two power-law function to the
ideal gas energy versus entropy curve and the good agreement
between the fitted and exact temperature, we apply the two
power-law fit function to the data for the strongly interacting
Fermi gas.

\begin{figure}
\centerline{\includegraphics[width=4.0in,clip]{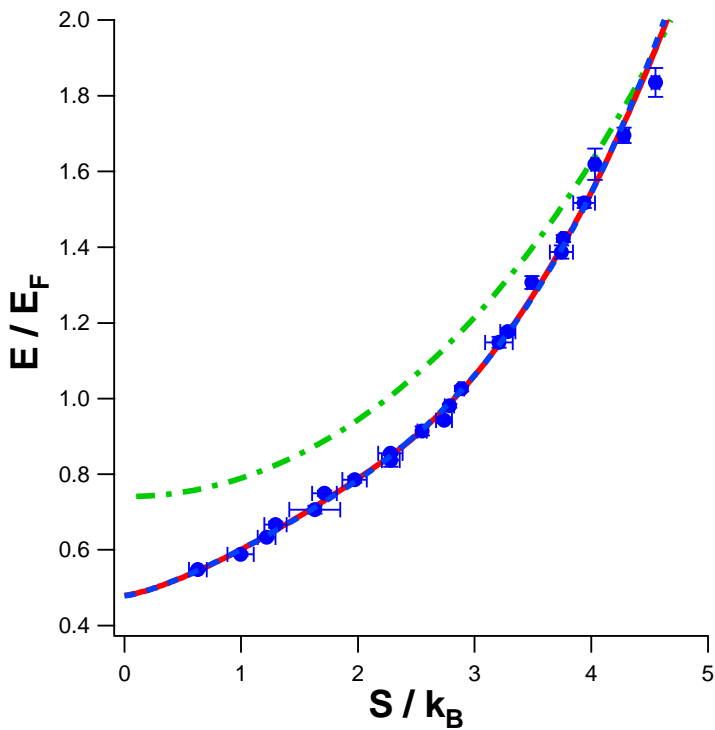}}
\caption {Color online. To determine the temperature, the
energy-entropy data are parameterized by joining two power-law fit
functions. The red solid line shows the fit that includes a heat
capacity jump, while the blue dashed curve shows the fit for a
continuous heat capacity (see \S~\ref{sec:temp}). For comparison,
the dot-dashed green curve shows $E(S)$ for an ideal Fermi gas.
For this figure, the ideal gas
approximation to the entropy is used, $S_{1200}/k_B$ of
Table~\ref{tbl:energyentropy}. \label{fig:energyentropyfit}}
\end{figure}

Fig.~\ref{fig:energyentropyfit} shows the fit (red solid curve)
obtained with a heat capacity jump  using
Eq.~\ref{eq:EvsSlowengtwopower} with $e=0$ and $a=0.12(1)$,
$b=1.35(11)$, $d=2.76(12)$, the ground state energy $E_0=0.48(1)$,
and the critical entropy $S_c=2.2(1)$. Also shown is the fit (blue
dashed curve) with continuous heat capacity ($e\neq 0$) and
$a=0.12(2)$, $b=1.31(17)$, $d=2.9(2)$, the ground state energy
$E_0=0.48(2)$, and the critical entropy $S_c=1.57(29)$.

\subsection{Estimating the Critical Parameters}

The fit functions for the $E(S)$ data for the strongly interacting
Fermi gas exhibit a significant change in the scaling of $E$ with
$S$  below and above $S_c$. The dramatic change in the power law
exponents for the strongly interacting gas suggests a transition
in the thermodynamic properties. The power law exponent is $2.9$
above $S_c$, comparable to that obtained for the ideal gas, where
$d=2.5$. The power law exponent below $S_c$ is $1.35$, which
corresponds to the low temperature dependence $E-E_0\propto
T^{3.86}$, close to that obtained in measurements of the heat
capacity, where the observed power law was $3.73$ after the
model-dependent calibration of the empirical
temperature~\cite{kinast05a}, see \S~\ref{sec:Ttwiddle}.

If we interpret $S_c$ as the critical entropy for a
superfluid-normal fluid transition in the strongly interacting
Fermi gas, then we can estimate the critical energy $E_c$ and the
critical temperature $T_c=(\partial E_<(S)/\partial S)_{S_c}$. For
the fits of Eq.~\ref{eq:EvsSlowengtwopower} with a heat capacity
jump $(e=0$) or with continuous heat capacity ($e\neq 0$), we
obtain
\begin{eqnarray}
E_c &=&E_0+a\,S_c^b\nonumber\\
T_c&=&ab\,S_c^{b-1}. \label{eq:criticalparam}
\end{eqnarray}
Using the fit parameters in Eq.~\ref{eq:criticalparam} yields
critical parameters of the strongly interacting Fermi gas, which
are summarized in Table~\ref{tbl:criticalparam}. The statistical
error estimates are from the fit, and do not include systematic
errors arising from the form of the fit function.

\begin{table}
\begin{center}
\begin{tabular}{|l|c|c|c|}  \hline
\rule[-3pt]{0pt}{14pt} & $S_c (k_B)$ & $E_c (E_F)$ & $T_c (T_F)$\\
\hline\hline \rule[-1pt]{0pt}{14pt} Expt $E(S)$ Fit$^1$& 2.2(1)  &
0.83(2) & 0.21(1) \\   \hline \rule[-1pt]{0pt}{14pt} Expt $E(S)$
Fit$^2$& 1.6(3)  & 0.70(5) & 0.185(15) \\   \hline
\rule[-1pt]{0pt}{14pt}Heat Capacity Experiment$^*$ &   & 0.85 &
0.20 \\   \hline \rule[-1pt]{0pt}{14pt}Theory Ref.~\cite{Torma} &
&  & 0.30 \\ \hline \rule[-1pt]{0pt}{14pt}Theory
Ref.~\cite{perali04} &  &  & 0.31 \\ \hline
\rule[-1pt]{0pt}{14pt}Theory Ref.~\cite{chen05thermodynamics} &  &
& 0.27 \\ \hline \rule[-1pt]{0pt}{14pt}Theory
Ref.~\cite{DrummondUniversal} &  &  & 0.29 \\   \hline
\rule[-1pt]{0pt}{14pt}Theory Ref.~\cite{BulgacThermodynamics} &
2.15 & 0.82 & 0.27 \\ \hline \rule[-1pt]{0pt}{14pt}Theory
Ref.~\cite{ZwergerUnitaryGas} & 1.61(5) & 0.667(10) & 0.214(7) \\
\hline
\end{tabular}
\end{center}
\caption{Critical parameters for a strongly interacting Fermi gas.
The experimental results are obtained from fits to the energy versus entropy
data using Eq.~\ref{eq:EvsSlowengtwopower}: Fit$^1$ uses $e=0$, and has a jump in heat capacity.
Fit$^2$ constrains $e$ so that there is no jump in heat capacity.
The theoretical results are presented for comparison. $^*$Using the present
experimental calibration of the measured empirical transition temperature,
see \S~\ref{sec:Ttwiddle}.\label{tbl:criticalparam}}
\end{table}

We  note that the fit function for $S(E)$ previously used in
Ref.~\cite{Luo07Entropy} to determine the temperature was
continuous in $S$ and $E$, but intentionally ignored the
continuous temperature constraint in order to determine the
entropy as a power of $E-E_0$ both above and below the joining
energy $E_c$. As the continuous temperature constraint is a
physical requirement, we consider the present estimate of the
temperature $T$ to be more useful for temperature calibrations and
for characterizing the physical properties of the gas than the
estimate of Ref.~\cite{Luo07Entropy}.

In contrast to the temperature $T$, the estimate of $T_c$ depends
on the value of the joining entropy $S_c$ that optimizes the fit
and is more sensitive to the form of fit function than the
temperature that is determined from the $E$ and $S$ data. For the
fit function $S(E)$ used in Ref.~\cite{Luo07Entropy}, the
temperatures determined by the fit function just above $E_c$,
$T_{c>}$, and below $E_c$, $T_{c<}$, were different. An average of
the slopes $1/T_{c>}$ and $1/T_{c<}$ was used to estimate the
critical temperature. From those fits, the critical energy was
found to be $E_c/E_F=0.94\pm 0.05$, the critical entropy per
particle was $S_c=2.7(\pm 0.2)\,k_B$. The estimated critical
temperature obtained from the average was
$T_c/T_F=0.29(+0.03/-0.02)$, significantly higher than than the
value $T_c/T_F=0.21$ obtained using
Eq.~\ref{eq:EvsSlowengtwopower}, which incorporates continuous
temperature.

We are able to substantiate the critical temperature
$T_c/T_F=0.21$ by using our data to experimentally calibrate the
temperature scales in two other experiments. In
\S~\ref{sec:onsetpaircond}, we find that this value is in very
good agreement with the estimate we obtain by calibrating the
ideal gas temperature observed for the onset of pair condensation.
Nearly the same transition temperature is obtained in
\S~\ref{sec:Ttwiddle} by using the $E(S)$ data to calibrate the
empirical transition temperature measured in heat capacity
experiments~\cite{kinast05a}.

Table~\ref{tbl:criticalparam} compares the critical parameters
estimated from the power-law fits to the  $E(S)$ data with the
predictions for a trapped unitary Fermi gas from several
theoretical groups. We note that calculations for a uniform
strongly interacting Fermi gas at unitarity~\cite{Burovski08}
yield a lower critical temperature, $T_c/T_F(n)=0.152(7)$, than
that of the trapped gas, where $T_F(n)$ is the Fermi temperature
corresponding to the uniform density $n$. Extrapolation of the
uniform gas critical temperature to that of the trapped gas shows
that the results are consistent~\cite{ZwergerUnitaryGas}.

\begin{figure}
\centerline{\includegraphics[width=
4.0in,clip]{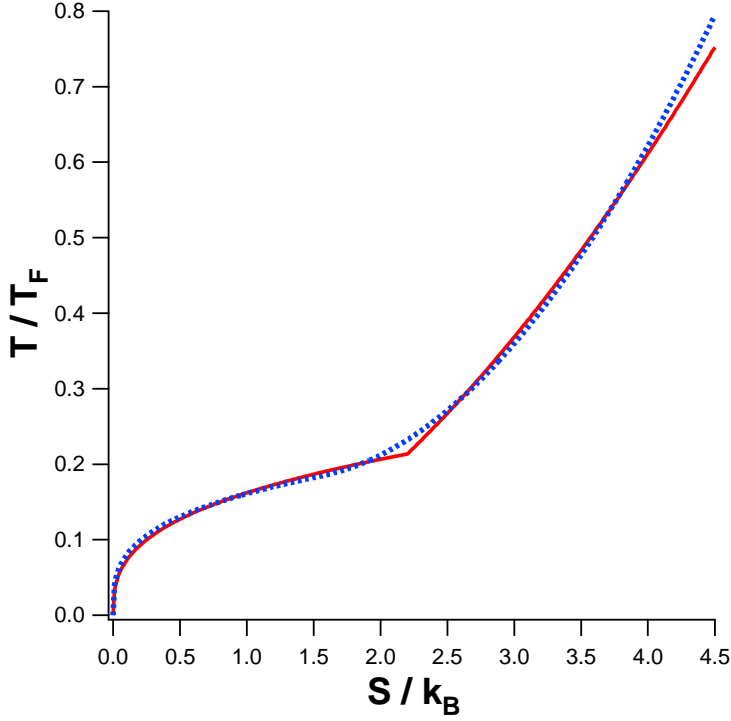}} \caption {Color online. The temperature versus
the entropy of a strongly interacting Fermi gas from the fits to the $E(S)$ measurement. The red solid curve is given by
Eq.~\ref{eq:TemperatureStrongInt} for $e=0$ (heat capacity jump)
and the blue dotted curve is for $e\neq 0$ (continuous heat
capacity).\label{fig:TempvsEntropy}}
\end{figure}

Using the parameters from the fits and
Eq.~\ref{eq:EvsSlowengtwopower}, the temperature of the strongly
interacting Fermi gas, in units of $T_F$ can be determined as a
function of the entropy per particle, in units of $k_B$,
\begin{eqnarray}
T_<(S)&=&T_c\,\left(\frac{S}{S_c}\right)^{b-1};
\,\,0\leq S\leq S_c\nonumber\\
T_>(S)&=&T_c\,\left(\frac{S}{S_c}\right)^{d-1}+\,2e(S-S_c);
\,\,S\geq S_c.
 \label{eq:TemperatureStrongInt}
\end{eqnarray}
Here $S_c$ is given in Table~\ref{tbl:criticalparam} from the
fits to the $E(S)$ data for the strongly interacting gas,
Eq.~\ref{eq:criticalparam} gives $T_c$.
Fig.~\ref{fig:TempvsEntropy} shows the temperature as a function
of entropy according to Eq.~\ref{eq:TemperatureStrongInt} for fits
with a heat capacity jump and for continuous heat capacity.

\section{Temperature Calibrations}

The estimates of the temperature of the strongly interacting Fermi
gas as a function of the entropy can be used to experimentally
calibrate the temperatures measured in other experiments, without
invoking any specific theoretical models.  The JILA group measures
the pair condensate fraction in  a strongly interacting Fermi gas
of $^{40}$K as a function of the initial temperature $T_{Ic}$ in
the noninteracting regime above the Feshbach
resonance~\cite{regal04,JinPhotoemission}. In these experiments, a
downward adiabatic sweep of the bias magnetic field to resonance
produces a strongly interacting sample. Using our $E(S)$ data, we
relate the endpoint temperatures for  adiabatic sweeps of the bias
magnetic field between the ideal and strongly interacting Fermi
gas regimes. We therefore obtain the critical temperature for the
onset of pair condensation in the strongly interacting Fermi gas,
and find very good agreement with our estimates based on
entropy-energy measurement.

In addition, we calibrate the empirical temperature based on the
cloud profiles, which was employed in our previous measurements of
the heat capacity~\cite{kinast05a}.

\subsection{Endpoint Temperature Calibration for Adiabatic Sweeps}
\label{sec:endpointtemp}

We relate the endpoint temperatures for an adiabatic sweep between
the strongly interacting and ideal Fermi gas regimes.
Eq.~\ref{eq:TemperatureStrongInt} gives the temperature of the
strongly interacting gas as a function of entropy, i.e., $T(S)$.

\begin{figure}
\centerline{\includegraphics[width= 4.0in,clip]{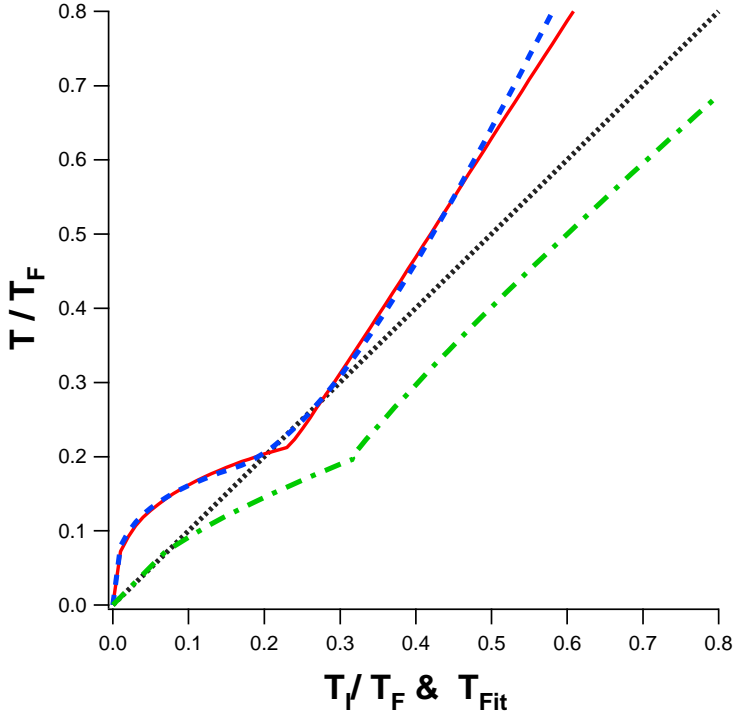}}
\caption {Color online. Experimental temperature calibrations:
Temperature of a strongly interacting Fermi gas $(T/T_F$) compared
to the temperature of a noninteracting Fermi gas ($T_I/T_F$) for
an adiabatic sweep between and strongly interacting and ideal
Fermi gas regimes (equal entropies). The solid red curve is
obtained from the fit to the $E(S)$ data with a heat capacity jump
and the dashed blue curve is obtained from the fit with continuous
heat capacity. The dashed green curve shows the value of the
empirical temperature $T_{fit}$, as obtained from the cloud
profiles in Ref.~\cite{kinast05a}, versus the corresponding
reduced temperature of the strongly interacting Fermi gas at the
same energy.  The dotted black line denotes equal temperatures.
\label{fig:tempSIvNI}}
\end{figure}

Next, we calculate the entropy per particle $S_I(T_I)$  for an
ideal Fermi gas in our gaussian trap, in units of $k_B$ , with
$T_I$ in units of $T_F$, as used in \S~\ref{sec:idealgas} to
determine $E_I(S_I)$. For an adiabatic sweep between the strongly
interacting and ideal Fermi gas regimes, where $S=S_I$, the
temperature of the strongly interacting gas is related to that for
the ideal Fermi gas by
\begin{equation}
T=T[S_I(T_I)], \label{eq:adiabactictempcal}
\end{equation}
which is shown in Fig.~\ref{fig:tempSIvNI}.\\

For an adiabatic sweep from the ideal Fermi gas regime to the
strongly interacting Fermi gas regime at low temperature $T<T_c$,
the reduced temperature of the strongly interacting gas  is
greater than or equal to that of the ideal gas. This arises
because the entropy of the strongly interacting gas  scales as a
higher power of the temperature than that of the ideal gas.

In our present experiments, we could not take data at high enough
temperatures to properly characterize the approach of the
temperature to the ideal gas regime. Above $T_c$,  our $E(S)$ data
are obtained over a limited range of energies $E\leq 2\,E_F$ to
avoid evaporation in our shallow trap. In this energy range, our
data are reasonably well fit by a  single power law. However, such
a power law fit cannot completely describe the higher temperature
regime.  We expect that the temperatures of the strongly
interacting gas and ideal gas must start to merge in the region $S
\cong 4\,k_B$, where the $E(S)$ data for the strongly interacting
gas nearly overlaps with the $E(S)$ curve for an ideal gas, as
shown in Fig.~\ref{fig:energyentropy}.

From Fig.~\ref{fig:TempvsEntropy}, $S>4\,k_B$ corresponds to
$T/T_F>0.6$, approximately the place where the calibrations from
the two different power law fits (for $e=0$ and $e\neq 0$) begin
to differ in Fig.~\ref{fig:tempSIvNI}. We therefore expect that
the single power law fit overestimates the temperature $T$ of the
strongly interacting gas for $T>0.6\,T_F$, yielding a trend away
from ideal gas temperature, in contrast to the expected merging at
high temperature.

\subsection{Critical Temperature for the Onset of Pair Condensation}
\label{sec:onsetpaircond}

In Ref.~\cite{regal04}, projection experiments measure the ideal
Fermi gas temperature $T_{Ic}$ where pair condensation  first
appears. In those experiments, $T_{Ic}$ is estimated to be
$0.18(2),T_F$~\cite{JinPhotoemission}. From the calibration,
Fig.~\ref{fig:tempSIvNI}, we see that for $T_{Ic}=0.18\,T_F$, the
corresponding temperature of the strongly interacting gas is
$T_c=0.19(2)\,T_F$ for both the red solid and blue dashed curves,
which is almost the same as the ideal gas value. The critical
temperature of the strongly interacting gas for the onset of pair
condensation is then $0.19(2)\,T_F$, in very good agreement with the
values $T_c=0.21(1)\,T_F$ and $T_c=0.185(15)\,T_F$ that we obtain
from the two fits to the $E(S)$ measurements. This substantiates
the conjecture that the change in the power law behavior observed
at $T_c$ in our experiments corresponds to the superfluid
transition.

\subsection{Calibrating the Empirical Temperature obtained from
the Cloud Profiles}
\label{sec:Ttwiddle}

In our previous study of the heat capacity, we determined an
empirical temperature $T_{fit}\equiv\tilde{T}$ as a function of
the total energy of the gas~\cite{kinast05a,kinast05b}. The gas
was initially cooled close to the ground state and a known energy
was added by a release and recapture method. Then a Thomas-Fermi
profile for an ideal Fermi gas was fit to the low temperature
cloud profiles to determine the Fermi radius. Holding the Fermi
radius constant, the best fit to the cloud profiles at higher
temperatures determined the effective reduced temperature, which
is denoted $\tilde{T}$. The $E(\tilde{T})$ data~\cite{kinast05a}
was observed to scale as $E-E_0=1.54\,E_F\,\tilde{T}^{1.43}$ for
$\tilde{T}\geq 0.33$, while below $\tilde{T}=0.33$, the energy was
found to scale as $E-E_0=4.9\,E_F\,\tilde{T}^{2.53}$. The
transition point occurs at an energy $E_c=0.85\,E_F$, which is
close to the value $0.83\,E_F$ obtained from power-law fit to the
$E(S)$ data for the fit with a heat capacity jump. Assuming that
$\tilde{T}_c=0.33$ corresponds to the superfluid-normal fluid
transition, we can determine the corresponding value of $T_c/T_F$
for the strongly interacting gas.

To calibrate the empirical temperature we start with $E(\tilde{T})$.
Then, as discussed in \S~\ref{sec:energyvstemp},
Eq.~\ref{eq:EvsSlowengtwopower} determines $E(T)$ and hence $T(E)$from the fits to
our $E(S)$ data.  Hence $T(\tilde{T})=T[E(\tilde{T})]$, where
$T\equiv T/T_F$ is the reduced temperature of the strongly interacting
gas and $E\equiv E/E_F$ is the reduced energy.  For simplicity,  we give the analytic results obtained using the $e=0$ fit to the $E(S)$ data,
\begin{eqnarray}
\frac{T}{T_F}&=&0.42\,\tilde{T}^{0.66};\,\,\,0\leq\tilde{T}\leq0.33\nonumber\\
\frac{T}{T_F}&=&\left(0.80\,\tilde{T}^{1.43}-0.09\right)^{0.64};\,\,\,\tilde{T}\geq
0.33. \label{eq:TildeTcalib}
\end{eqnarray}
 Fig.~\ref{fig:tempSIvNI} shows the
full calibration (green dashed curve). For comparison, the
calibration obtained from the pseudogap theory of the cloud
profiles gave $T/T_F=0.54\,\tilde{T}^{0.67}$ for $\tilde{T}\leq
0.33$, and $T/T_F=0.71\,\tilde{T}$ above $\tilde{T}=0.33$. For
$\tilde{T}_c=0.33$, we obtain from Eq.~\ref{eq:TildeTcalib}
$T_c/T_F=0.20$ (see Fig.~\ref{fig:tempSIvNI}), in good agreement
with the value obtained for the onset of pair condensation and
with the values $T_c=0.185(15)$ and $T_c=0.21(1)$ determined from the fits to the
$E(S)$ data.

\section{Measuring the Ground State Energy}

Measurement of the ground state energy of a unitary Fermi gas
provides a stringent test of competing many-body theoretical
predictions and is therefore of great interest. For a unitary
Fermi gas of uniform density in a 50-50 mixture of two spin
states, the ground state energy per particle can be written as
\begin{equation}
E_g=(1+\beta)\,\frac{3}{5}\epsilon_F(n), \label{eq:betadefinition}
\end{equation}
where $\epsilon_F(n)$ is the local Fermi energy corresponding to
the density $n$. The ground state energy of the unitary Fermi gas
differs by a universal factor $\xi\equiv 1+\beta$ from that of an
ideal Fermi gas at the same density. The precise value of $\xi$
has been of particular interest in the context of neutron
matter~\cite{Bertsch,Baker,Heiselberg,carlson03}, and can be
measured in unitary Fermi gas experiments~\cite{ohara02,gehm03}.

The sound speed  at temperatures near the ground state determines
$\beta$ according to Eq.~\ref{eq:soundres}. We have made precision
measurements of the sound speed in a trapped Fermi gas at the
Feshbach resonance~\cite{Joseph07Sound}. At 834 G, we vary the
density by a factor of 30 to demonstrate universal scaling and
obtain the value $c_0/ \textrm{v}_F=0.362(6)$. Using
Eq.~\ref{eq:soundres} then yields $\beta =-0.565(15)$. Note that
the reference Fermi velocity $\textrm{v}_F$ depends on the Fermi
energy of an ideal gas at the trap center and hence on both the
trap frequencies and atom number (as $N^{1/6}$), which are
carefully measured to minimize systematic
errors~\cite{Joseph07Sound}. While the energy of the gas as
measured from the mean square cloud size was close to the ground
state value, the precise temperature of the gas was not
determined.

The universal parameter $\beta$ also can be determined by
measuring the ground state energy $E_0$ of a harmonically trapped
unitary Fermi gas, which is given by
\begin{equation}
E_0=\frac{3}{4}\sqrt{1+\beta}\,E_F. \label{eq:HObeta}
\end{equation}
Our $E(S)$ data enables a new determination of $E_0$ by
extrapolating the measured energy $E(S)$ to $S=0$. As pointed out
by Hui et al.~\cite{DrummondUniversal}, this method avoids a
systematic error arising when the finite temperature is not
determined in the measurements.

From both of our fit functions below $S_c$, we obtain
$E_{0}/E_F=0.48(1)$. Eq.~\ref{eq:HObeta} yields $\beta=-0.59(2)$.
This result is slightly more negative than that obtained in the
sound speed experiments, which is reasonable since the sound speed
measurements are done at finite temperature. Both results are in
very good agreement.

One possible systematic error in these measurements arises from
the determination of the atom number. The measurements of $\beta$
from the sound speed and from the energy-entropy measurements were
done in different laboratories. The close agreement is gratifying,
considering that the imaging systems that determine the atom
number employed $\sigma_-$-polarized light for the sound speed
experiments, while the entropy-energy measurements used
$x$-polarized light, for which the resonant optical cross section
is a factor of two smaller than for $\sigma_-$ polarization. To
examine the systematic error arising from the atom number
determination, we employ a third method to measure $\beta$ based
on the measured ratio of the cloud size at 840 G and at 1200 G,
which is number independent.

The ratio of the ground state mean square sizes for the weakly and
strongly interacting gases is predicted to be
\begin{equation}
r_0=\frac{\langle z^2\rangle_{0,1200}/z_F^2(1200)}{\langle
z^2\rangle_{0,840}/z_F^2(840)}=\frac{0.69}{(3/4)\sqrt{1+\beta}}.
\label{eq:groundstateratio}
\end{equation}
Note that we obtain $\langle z^2\rangle_{0,1200}/z_{F1200}^2=0.69$
from a mean field calculation~\cite{luothesis},  in agreement with
that obtained using a many-body
calculation~\cite{DrummondUniversal,BulgacThermodynamics}.

Our measurements for the ground state mean square size at 1200 G
are accomplished by fitting a Sommerfeld expansion of the axial
density for an ideal Fermi gas to the cloud profile~\cite{Luo07Entropy,gehm03}.
The fit determines the Fermi radius $\sigma_z$ and reduced
temperature $T/T_F$, yielding $\langle
z^2\rangle_{0,1200}=\sigma_z^2/8=0.71\,z_F^2(1200)$ for $T=0$,
close to the predicted value of $0.69$. The ground state energy
$E_0=0.48\,E_F(840)$ at 840 G from the entropy-energy experiments
determines the ground state mean square size as $\langle
z^2\rangle_{0,840}=0.48\,z_F^2(840)$. Hence, $r_0=0.71/0.48=1.48$.
The corresponding $\beta=-0.61(2)$ from
Eq.~\ref{eq:groundstateratio}. Since the mean square sizes are
determined from the images and the ratio  $z_F^2(840)/z_F^2(1200)$
is number independent, this result shows that the systematic error
arising from the number measurement is within the quoted error
estimate.

We also can determine $\beta$  by directly extrapolating to zero
entropy the ratio of the axial mean square size of the weakly
interacting Fermi gas at 1200 G to that of strongly interacting
gas at 840 G.  When this is done, we obtain
$\beta =-0.58$, in very
good agreement with the estimates based on the sound speed and
ground state energy.

Finally, we can estimate the correction to the ground state energy
arising from the finite scattering length at 840 G,
$a_S=-73616\,a_0$. For the trap conditions in the $E(S)$
measurements, $k_Fa_S=-18$, where $k_F$ is the wavevector for an
ideal Fermi gas at the trap center. To estimate the true unitary
ground state energy at $a_S=\infty$, we first determine the
leading order $1/(a_S\,k_F(n))$ correction to the trapped atom
density, where $k_F(n)$ is the local Fermi wavevector
corresponding to the density $n$. The local chemical potential is
estimated from Ref.~\cite{ChinSimpleMF}. Using  the notation of
Eq.~\ref{eq:reducedTruePotential} and a harmonic approximation,
the corrected density yields  $\langle r^2/\sigma^2\rangle =
(3/8)[1-(128/105\pi)(0.64/k_Fa_S)]$, where $\sigma$ is the Fermi
radius for the unitary gas. According to the virial theorem (see
Eq.~\ref{eq:energymeas}), the mean square size and energy of the
unitary gas are corrected by the same factor. The unitary ground
state energy is then
\begin{equation}
E_0(\infty)=\frac{E_0(k_Fa_S)}{1-(128/105\pi)(0.64/k_Fa_S)}.
\label{eq:energycorrection}
\end{equation}
For $k_Fa_s=-18$, we obtain $E_0(\infty)=0.986\,E_0(-18)$ and the
value of $\beta = -0.59(2)$ obtained directly from
$E(S=0)=E_0=0.48(1)$ is shifted to $\beta =-0.60(2)$. We also
obtain the corrected value of $r_0=0.71/(0.986*0.48)=1.50$ in
Eq.~\ref{eq:groundstateratio} and $\beta =-0.62(2)$.

Table~\ref{tbl:beta} compares the values of $\beta$ obtained in
our experiments to several recent predictions. Note that the table
does  not include the finite $k_Fa$ correction for the $E(S)$
measurement at 840 G described above.

\begin{table}
\begin{center}
\begin{tabular}{|l|c|c|c|}  \hline
\rule[-3pt]{0pt}{14pt} & $\beta$ \\  \hline \hline
\rule[-1pt]{0pt}{14pt}$E(S)$ Experiment & -0.59(2) \\ \hline
\rule[-1pt]{0pt}{14pt}Sound Velocity Experiment &-0.565(15)\\
\hline
\rule[-1pt]{0pt}{14pt}Cloud Size Ratio Experiment &-0.61(2)\\
\hline
\rule[-1pt]{0pt}{14pt}Ref.~\cite{carlson03,chang04,Astrakharchik}&
-0.58(1)\\ \hline \rule[-1pt]{0pt}{14pt}Ref.~\cite{BulgacUnitary}
&$-0.56(3)$\\ \hline
\rule[-1pt]{0pt}{14pt}Ref.~\cite{DrummondUniversal} & -0.599 \\
\hline \rule[-1pt]{0pt}{14pt}Ref.~\cite{carlson08}& -0.60(1)\\
\hline
\rule[-1pt]{0pt}{14pt}Ref.~\cite{ZwergerUnitaryGas} & -0.646(4)\\
\hline
\end{tabular}
\end{center}
\caption{Universal interaction parameter $\beta$.}
\label{tbl:beta}
\end{table}

\section{Universal Thermodynamic Functions}

Using the  $E(S)$ data for the strongly interacting Fermi gas and
the temperature determined from the two power-law fits, we
estimate several universal functions. First, we determine the
dependence of the energy on temperature $E(T)$ and the
corresponding heat capacity, $C(T)$. Then we find the global
chemical potential of the trapped gas as a function of the energy
$\mu_g(E)$.

\subsection{Energy versus Temperature}
\label{sec:energyvstemp}

\begin{figure}
\centerline{\includegraphics[width= 4.0in,clip]{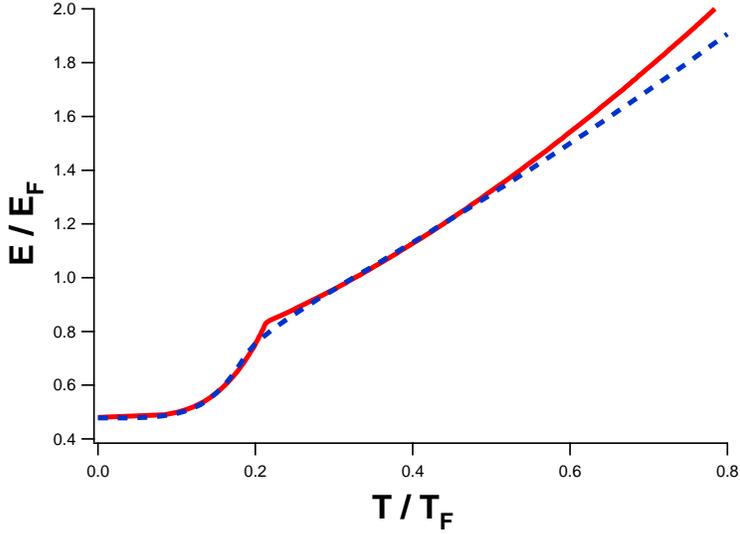}}
\caption {Color online. The energy of a strongly interacting Fermi gas versus
temperature, from the fits to the $E(S)$ data. The red curve shows
$E(T)$ as determined from the fit with a heat capacity jump
($e=0$) in Eq.~\ref{eq:EvsSlowengtwopower}. The blue dashed curve
shows $E(T)$ as determined from the fit with continuous heat
capacity ($e\neq 0$).\label{fig:EnergyvsTemp}}
\end{figure}

The energy is readily determined as a function of temperature
using Eq.~\ref{eq:EvsSlowengtwopower} for the case where there is a heat capacity jump and $e=0$,
\begin{eqnarray}
E_<(T)&=&E_c+{\frac{S_cT_c}{b}}\left[\left(\frac{T}{T_c}\right)^{\frac{b}{b-1}}-1\right];\,\,\,0\leq T\leq T_c\nonumber\\
E_>(T)&=&E_c+{\frac{S_cT_c}{d}}\left[\left(\frac{T}{T_c}\right)^{\frac{d}{d-1}}-1\right];
\,\,\,T\geq T_c , \label{eq:EnergyvsTemp}
\end{eqnarray}
where the energy (temperature) is given in units of $E_F$ ($T_F$)
and the critical energy $E_c$ is
\begin{equation}
E_c=E_0+S_cT_c/b, \label{eq:criticalenergy}
\end{equation}
with $E_0$ the ground state energy. For the case with $e\neq 0$,
where the heat capacity is continuous, we determine the ordered
pairs $[E(S),T(S)]$ as a function of $S$ and plot $E(T)$.
Fig.~\ref{fig:EnergyvsTemp} shows the results using the best fits for both
cases.

Of particular interest is the low temperature power law. For
$e=0$, we obtain $b=1.35$ and $b/(b-1)=3.86$. Since $b$ is near
$4/3$, the energy relative to the ground state scales
approximately as $T^4$. This is consistent with sound modes
dominating the low energy excitations. However, one would expect
instead that the free fermions on the edges of the trapped cloud
would make an important contribution to the low energy
excitations~\cite{chen05thermodynamics}. Over an extended range of
$T<T_c$, the net entropy arising from the Bose and Fermi
excitations has been predicted to scale as $T^2$, yielding an
energy scaling~\cite{chen05thermodynamics} as $E-E_0\propto T^3$.
In this case, one would expect that $E-E_0\propto S^{3/2}$, i.e.,
$b=3/2$ in Eq.~\ref{eq:EnergyvsTemp}, so that $b/(b-1)=3$. Hence,
the low energy  power law exponents for the entropy should be
between $4/3$ and $3/2$, which is barely distinguishable for our
data.

\subsection{Heat Capacity versus Temperature}

\begin{figure}
\centerline{\includegraphics[width= 4.0in,clip]{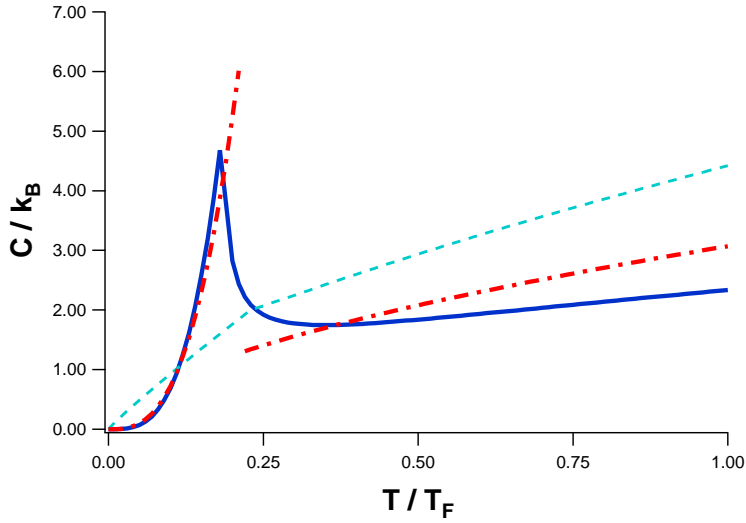}}
\caption {Color online. Heat capacity versus temperature given by
Eq.~\ref{eq:heatcapacity} for a strongly interacting Fermi gas.
The red dot-dashed curve shows the heat capacity when there is a
jump at $T_c/T_F=0.21$. The blue solid curve shows the heat
capacity when the heat capacity is continuous. For comparison, the
light-blue dashed curve shows the heat capacity obtained for an
ideal Fermi gas (using the fit function of
Fig.~\ref{fig:energyentropyfit}).\label{fig:CvsT}}
\end{figure}

The heat capacity at constant trap depth $C=dE/dT$ is readily
obtained from Eq.~\ref{eq:EnergyvsTemp} (where there is a heat
capacity jump, since we have constrained $e=0$ in
Eq.~\ref{eq:EvsSlowengtwopower}). For this parameterization,
\begin{eqnarray}
C_<(T)&=& \frac{S_c}{b-1}\left(\frac{T}{T_c}\right)^{\frac{1}{b-1}};\,\,\,0\leq T\leq T_c\nonumber\\
C_>(T)&=&\frac{S_c}{d-1}\left(\frac{T}{T_c}\right)^{\frac{1}{d-1}};\,\,\,T\geq
T_c , \label{eq:heatcapacity}
\end{eqnarray}
where $T$ and $T_c$ are given in units of $T_F$, and $S_c$ is
given in units of $k_B$. For the fit with a continuous heat
capacity, we use $T(S)$ to find $C(S)=T(S)/(dT/dS)$, and plot the
ordered pairs $[C(S),T(S)]$.  The heat capacity curves for both
cases are shown in Fig.~\ref{fig:CvsT}.

According to Eq.~\ref{eq:heatcapacity},  a jump in heat capacity
occurs at $S_c$: $C_<(T_c)=S_c/(b-1)$ and $C_>(T_c)=S_c/(d-1)$
differ when the power law exponents $b$ and $d$ are different.
This is a consequence of the simple two power-law structure
assumed for the fit function $E(S)$ given by
Eq.~\ref{eq:EvsSlowengtwopower} for $e=0$, and cannot be taken as
proof of a true heat capacity jump.  At present, the precise
nature of the behavior near the critical temperature cannot be
determined from our data, and it remains an open question whether
the data exhibits a heat capacity jump or a continuous heat
capacity.

\subsection{Global Chemical Potential versus Energy}

The global chemical potential $\mu_g$ is readily determined from
the fits to the $E(S)$ data for a strongly interacting Fermi gas,
which obeys universal thermodynamics. The local energy density
generally takes the form $\varepsilon=T\,s+ \mu\,n - P$, where
$\varepsilon$ is the local internal energy, which includes the
kinetic energy and the interaction energy. Here, $n$ is the local
density, $\mu$ is the local chemical potential, $P$ is the
pressure and $s$ is the total entropy per unit volume.

The local chemical potential can be written as $\mu = \mu_g-U$,
where $U$ is the trap potential. In the universal regime, where
the local pressure depends only on the local density and
temperature, we have $P=2\,\varepsilon/3$, as noted by
Ho~\cite{ho04}. Hence, $5\varepsilon/3=T\,s+(\mu_g-U)\,n$.
Integrating both sides over the trap volume and using $\int
d^3\mathbf{x}\,\varepsilon= N\,E-N\langle U\rangle$, where $E$ and
$\langle U\rangle$) are the total energy and average potential
energy per particle, respectively, we obtain
\begin{equation}
\frac{5}{3}NE-\frac{2}{3}N\langle U\rangle = N\,TS\,+\,\mu_g\,N,
\label{eq:globalchempot1}
\end{equation}
where $S$ is the entropy per particle.  For simplicity, we assume
harmonic confinement and use the virial theorem result, $\langle
U\rangle = E/2$,  from Eq.~\ref{eq:energymeas}, which holds in the
universal regime. Then, Eq.~\ref{eq:globalchempot1} yields the
global chemical potential of a harmonically trapped Fermi gas in
the universal regime,
\begin{equation}
\mu_g=\frac{4}{3}E-TS. \label{eq:globalchempot}
\end{equation}

By using the fit to the measured entropy-energy data to obtain the
temperature $T=\partial E/\partial S$ from
Eq.~\ref{eq:EvsSlowengtwopower}, the global chemical potential of
a trapped unitary Fermi gas can be calculated from
Eq.~\ref{eq:globalchempot}. For $e=0$, where the heat capacity has
a jump, the simple power law fits above and below $E_c$ each yield
a different linear dependence of $\mu_g$ on $E$,
\begin{eqnarray}
\mu_g(E)&=&\frac{4}{3}E_0+\left(\frac{4}{3}-b\right)(E-E_0);\,\,\,E_0\leq
E\leq E_c\nonumber\\
\mu_g(E)&=&\mu_g(E_c)+\left(\frac{4}{3}-d\right)(E-E_c);\,\,\,E\geq
E_c, \label{eq:globalchempotvsE}
\end{eqnarray}
where $\mu_g(E_c)=4E_0/3+(4/3-b)(E_c-E_0)$.

\begin{figure}
\centerline{\includegraphics[width= 4.0in,clip]{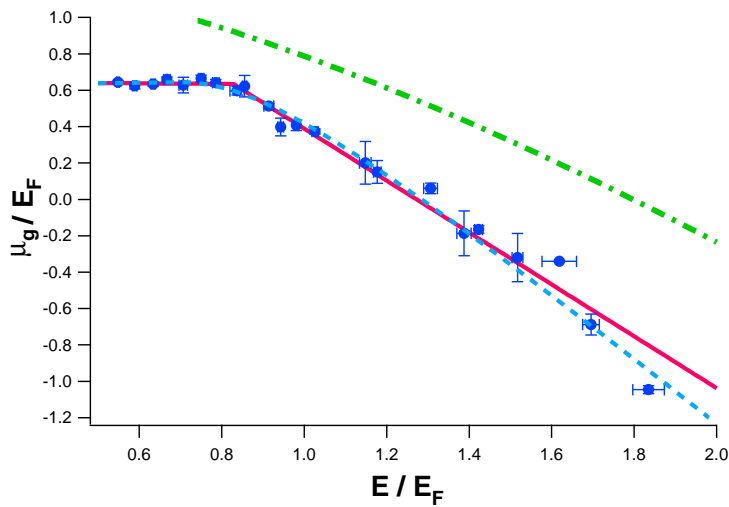}}
\caption {Color online. The global chemical potential versus the total energy of
a strongly interacting Fermi gas. The data points are calculated
from the measured $E-S$ data and the fitted $T$, where $T$ is
determined by the fit parameters in Eq.~\ref{eq:EvsSlowengtwopower} for $e=0$.
The standard deviation for each point of the chemical potential is
determined by the standard deviation of the measured $E-S$ data.
The solid red lines (heat capacity jump) and blue dashed curve
(heat capacity continuous) are determined by the fit parameters
used in Eq.~\ref{eq:EvsSlowengtwopower} according to
Eq.~\ref{eq:globalchempotvsE}. The green dot-dashed curve shows
the  ideal Fermi gas result for the fit function of
Fig.~\ref{fig:energyentropyNI2PFit}.\label{fig:chemenergy}}
\end{figure}

We plot the chemical potential in Fig.~\ref{fig:chemenergy}. The
data points are obtained using Eq.~\ref{eq:globalchempot} with the
measured energy $E$ and entropy $S$ and the temperature determined
from the  fit to the $E(S)$ data, using $e=0$ in
Eq.~\ref{eq:EvsSlowengtwopower}, i.e., with a heat capacity jump.
The solid red curve is given by Eq.~\ref{eq:globalchempotvsE}. We
note that the low temperature data points in $E(S)$ are best fit
with the power law $b=1.35$, which is close to $4/3$. According to
Eq.~\ref{eq:globalchempotvsE}, this produces a slope near zero for
$E_0\leq E\leq E_c$. Since the power-law fit above $E_c$ gives
$d=2.76$, the slope according to Eq.~\ref{eq:globalchempotvsE}
changes from nearly zero for $E_0\leq E\leq E_c$ to negative for
$E\geq E_c$.

Note that from Eq.~\ref{eq:globalchempot}, we obtain the slope
\begin{equation}
\frac{\partial \mu_g}{\partial E}=\frac{1}{3}-\frac{S}{C}\,.
\label{eq:slopeglobalchempot}
\end{equation}
Since the entropy $S$ is continuous, we see that a jump in the
heat capacity produces a corresponding jump in the slope of
$\mu_g$ versus $E$.

For comparison, Fig.~\ref{fig:chemenergy} also shows the chemical
potential obtained for $e\neq 0$ in
Eq.~\ref{eq:EvsSlowengtwopower}, where the heat capacity is
continuous (blue dashed curve).

\section{Conclusion}
We have studied the thermodynamic properties of a strongly
interacting Fermi gas by measuring both the energy and the
entropy. The model-independent data obtained in both the superfluid
and the normal fluid regimes do not employ any specific theoretical
calibrations, and therefore can be used as a benchmark to test the
predictions from many-body theories and simulations.
Parameterizing the energy-entropy data determines the temperature
of  the strongly interacting Fermi gas and also yields estimates
of the critical parameters. We use the measured data to calibrate
two different temperature scales that were employed in
observations of the onset of pair condensation and in heat
capacity studies. These calibrations yield  critical temperatures
in good agreement with the results estimated from our
energy-entropy data. Our data does not determine whether the heat
capacity exhibits a jump or is continuous at the critical
temperature. However, for a finite system with nonuniform density,
the latter is most likely. Considering that there is huge interest
in determining the detailed behavior of the superfluid transition
in a strongly interacting Fermi gas~\cite{ZwergerBECBCS}, more
precise determinations of the critical temperature, the heat
capacity, and the chemical potential near the critical point, as
well as the high temperature behavior and the approach to the
ideal gas limit, will be important topics for future research.

\section*{Acknowledgements}
This research has been supported by the Physics divisions of the
Army Research Office and the National Science Foundation, the
Physics for Exploration program of the National Aeronautics and
Space Administration, and the Chemical Sciences, Geosciences and
Biosciences Division of the Office of Basic Energy Sciences,
Office of Science, U. S. Department of Energy. We thank Willie Ong
for a careful reading of the manuscript.


\end{document}